\documentclass[aip,jcp,reprint,floatfix,a4paper,amsmath]{revtex4-1}

\usepackage{graphicx}
\usepackage[top=18mm,left=18mm,right=18mm,bottom=18mm]{geometry}

\begin{document}

\title{Determination of hydrogen cluster velocities and comparison with numerical calculations}

\author{A. T\"{a}schner}
\email[Author to whom correspondence should be addressed. Electronic mail: ]{taschna@uni-muenster.de}
\author{E. K\"{o}hler}
\author{H.-W.~Ortjohann}
\author{A. Khoukaz}
\affiliation{Institut f\"{u}r Kernphysik, Westf\"{a}lische
  Wilhelms-Universit\"{a}t M\"{u}nster, D-48149 M\"{u}nster, Germany}

\date{\today}

\begin{abstract}
The use of powerful hydrogen cluster jet targets in storage ring experiments
led to the need of precise data on the mean cluster velocity as
function of the stagnation temperature and pressure for the determination of
the volume density of the target beams.
For this purpose a large data set of hydrogen cluster velocity distributions
and mean velocities was measured at a high density hydrogen cluster jet target
using a trumpet shaped nozzle. The measurements have been performed at pressures
above and below the critical pressure and for a broad range of temperatures
relevant for target operation, e.g., at storage ring experiments. The used
experimental method is described which allows for the velocity measurement
of single clusters using a time-of-flight technique.
Since this method is rather time-consuming and these measurements are typically
interfering negatively with storage ring experiments, a method for a precise
calculation of these mean velocities was needed. For this, the determined
mean cluster velocities are compared with model calculations based on an
isentropic one-dimensional van der Waals gas.
Based on the obtained data and the presented numerical calculations, a new method
has been developed which allows to predict the mean cluster velocities with an
accuracy of about 5\%. For this two cut-off parameters defining positions inside
the nozzle are introduced, which can be determined for a given nozzle by only
two velocity measurements.
\end{abstract}


\maketitle

\section{Introduction}

Cluster beams have been studied in the last century extensively
with respect to their physical and chemical properties and even today the interest in
technological applications is increasing rapidly\cite{Pauly2000}. Prominent
examples are the use for cluster beam deposition, cluster impact lithography,
and the application as target beams in, e.g., storage ring experiments which
has started only in the last few decades.
For applications in hadron physics experiments or in experiments with high intense
laser beams, the most important advantage is that they provide a pure target
material inside a vacuum chamber with densities in the range between gas beams
and the solid state targets.
Cluster beams consisting of particles with sizes from the nanometer
to the micrometer scale propagate through vacuum with almost no increase of
their angular spread, so that it is possible to provide a spatially well
defined interaction zone for, e.g., a particle beam in a storage ring or
a laser beam. 

In hadron physics experiments at a storage ring the cluster beams are typically produced 
by expansion of gaseous materials in Laval type nozzles. Although such targets
can be operated in principal with all kind of gaseous materials ranging from hydrogen
to, e.g., xenon, the use of hydrogen is of special interest as effective proton
target for the investigation of elementary reactions. 
Examples for experimental facilities using such hydrogen cluster beams as 
internal targets at a storage ring are the ANKE\cite{Barsov2001} experiment and the 
former \mbox{COSY-11}\cite{Brauksiepe1996} experiment, both 
situated at the COSY\cite{Maier1997} accelerator of the Forschungszentrum J\"{u}lich.
For the established internal target experiments mainly the density and the purity
of the cluster beams were important, but for the design of new experimental facilities
which can be operated with event rates increased by one or two orders
of magnitude, the precise knowledge of microscopic properties, like velocity or
mass distributions, or time structure, became of high importance.
This information is especially important for the simulation of the interaction
between intense pulsed laser beams and cluster beams.
An example for a future experiment at a storage ring is the $4\pi$ detector
PANDA\cite{PANDA2005} at the planned accelerator centrum FAIR in Darmstadt (Germany). For this
experiment a cluster jet target has been developed at the University of M\"{u}nster
where the number of target atoms per unit area is above
$10^{15}\,\text{atoms/cm}^2$ at a distance of $2.1\,\text{m}$ from the
nozzle.
A detailed description of this prototype is presented in
Ref.~\onlinecite{Taeschner2011} and Ref.~\onlinecite{Taeschner2013}.
The measurements presented in this work have been performed at this prototype. 

\begin{figure}
\includegraphics[width=\columnwidth]{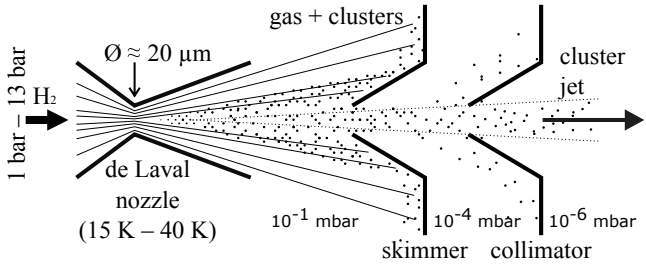}
\caption{\label{fig:PrincipleOfOperation}Schematic view of the principle of
operation of a cluster jet source which produces clusters from
hydrogen gas in front of the Laval nozzle.}
\end{figure}

In these targets the clusters are produced from ultra-clean cold compressed hydrogen fluid
with temperatures of, e.g., $25\,\text{K}$ and pressures of
about $18\,\text{bar}$, which is pressed through a Laval
nozzle with a minimum diameter in the order of $20\,\mu\text{m}$. During the
expansion of the fluid through the nozzle into a first vacuum chamber clusters 
are produced. Directly behind the nozzle the shape of the jet beam, consisting of 
both clusters and a gas beam, is determined by the shape of the divergent part of 
the production nozzle. In order to prepare a well defined cluster beam for
experiments and to suppress the disturbing residual gas background from the 
gas beam, a set of two skimmers is used to separate differential pumping stages. 
The second skimmer, which is denoted in the following as collimator, determines
the final shape and size of the cluster beam at all further vacuum stages,
and especially at the interaction point with the beam of the storage ring 
in the scattering chamber. For more details see Ref.~\onlinecite{Taeschner2011}.
A schematic sketch of this setup operated with gaseous
hydrogen is shown in Fig.~\ref{fig:PrincipleOfOperation}. 

\begin{figure}
  \includegraphics[width=\columnwidth]{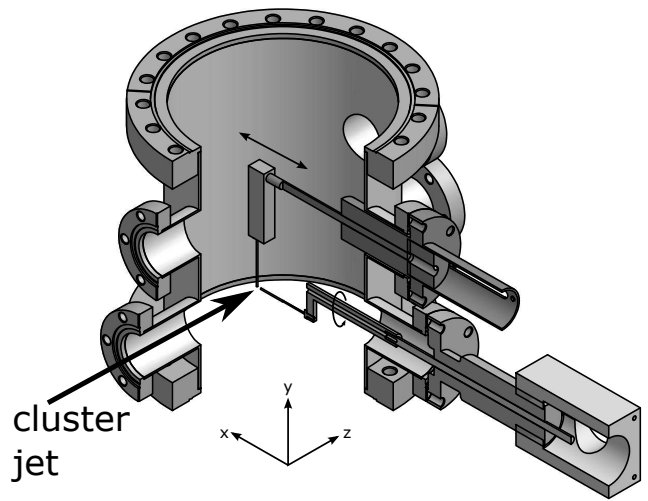}
  \caption{\label{fig:ScatteringChamber}Scattering chamber with moveable rods
	used to measure the target thickness.}
\end{figure}

\begin{figure}[b!]
  \includegraphics[width=\columnwidth]{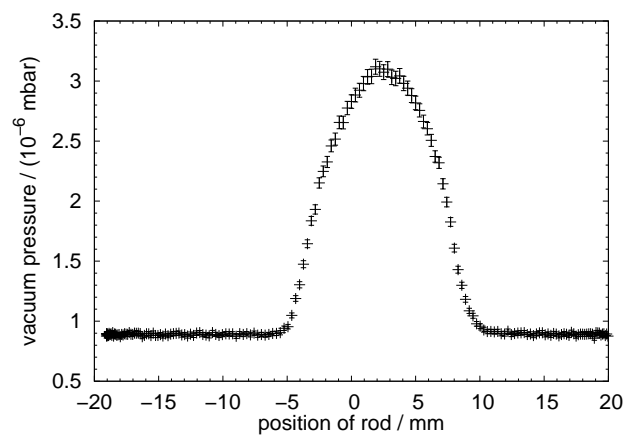}
  \caption{\label{fig:PressureProfile}Example of a pressure profile measured
	in the scattering chamber.}  
\end{figure}

For an optimized use as internal target in storage ring experiments, e.g., for 
hadron physics experiments, it is essential to determine and to adjust the
target thickness $n_{T}$,
the number of target atoms per unit area. With the knowledge of the thickness the
luminosity $L=f\,N_{C}\,n_{T}$ of the internal experiment can be calculated
(see, e.g.,\ Ref.~\onlinecite{Hinterberger2008}), where $f$ is the revolution frequency
and $N_{C}$ the number of circulating particles in the storage ring. Given the cross section
$\sigma$ of a specific reaction between target and storage ring beam particles, the
mean rate $\dot{N}_{R} = \sigma\,L$ of these reactions can be determined. This determination
is especially important for adjusting the target thickness to achieve a desired
reaction rate and to estimate the rate of background events.
Using a Cartesian coordinate system, where the cluster beam propagates
along the z axis and the stored beam along the x axis and assuming that the
transverse width of the stored beam is negligible compared to the size of the cluster beam, it is
possible to calculate the target thickness $n_{T}$ in units of number of atoms
per square centimeter at a specific distance $z_{0}$ behind the nozzle
directly from the volume density distribution~$\rho(x,y,z_{0})$
\cite{Taeschner2011}:
\begin{equation}
n_{T} = \frac{N_{A}}{M_{a}}\int_{-\infty}^{+\infty}{\rho(x,y,z_{0})\,\text{d}x}\,,
\end{equation}
where $M_{a}$ is the molar mass of the gas atoms and $N_{A}$ the Avogadro constant.
In case of the cluster jet target the thickness can be measured by
inserting movable rods into the cluster jet. At the cluster target prototype for the
PANDA experiment such rods are mounted in a scattering chamber located approximately 
two meters behind the nozzle (Fig.~\ref{fig:ScatteringChamber}). 
Here the rod diameter of $d=1\,\text{mm}$ is small compared to the size of the cluster~jet
which is typically in the order of about $10\,\text{mm}$. Clusters colliding with these rods
are stopped and lead by evaporation to an increase of the vacuum pressure
in this chamber. In Fig.~\ref{fig:PressureProfile} a typical measurement
of the vacuum pressure is presented, where the pressure increase is
plotted as a function of the rod position. With such kind of measurements the 
size as well as the position of the cluster jet within the vacuum stage can be determined
easily. Moreover, assuming a specific volume density distribution $\rho(x,y,z_{0})$ this pressure
profile can be described by the following equation \cite{Taeschner2011}:
  \begin{equation}
    p(x) = p_\text{b} + \frac{u\,R\,T}{S\,M}\int\limits_{x-x_0-d/2}^{x-x_0+d/2}\text{d}x'
           \int\limits_{-\infty}^{+\infty}\text{d}y'\,\rho(x',y',z_{0}).
  \end{equation}
In this equation $p_\text{b}$	is the background pressure, $x_{0}$~the center
position of the cluster jet, $R$~the universal gas constant, $M$~the molar mass
of the gas, $S$~the known pumping speed of the used pumping system, and $u$ the 
mean velocity of the clusters.
Note that the velocity $u$ depends on the stagnation condition, i.e., the
temperature $T_{0}$ and the pressure $p_{0}$, as well as on the nozzle geometry.
However, for constant numbers for $p$ and $T$ the velocity $u$ is constant.
Therefore, in order to calculate the target thickness in a first step the
target density distribution~$\rho(x',y',z_{0})$ has to be adjusted to describe
the relative shape of the measured pressure profile. For an absolute target
density determination the mean velocity $u$ has to be known.
In order to minimize the
uncertainty of the extracted volume density, the mean velocity has to be
measured with an accuracy which does not exceed the uncertainty of the pressure
and of the pumping speed. At the presented setup both uncertainties amount
to about~10\%.

Since the cluster beams from the described cluster target sources
are optimized for highest volume densities, the used nozzles differ
significantly both in shape and size from the ones commonly used by groups specialized
in the investigation of velocity and mass distributions, e.g., work on hydrogen
clusters reported in Ref.~\onlinecite{Knuth1995}. In the cited work a pin hole
nozzle with a minimum diameter of $5\,\mu\text{m}$ was used, whereas for
the production of the cluster beams described here trumped shaped nozzle
geometries with minimum diameters of about $20\,\mu\text{m}$ are used.
Both factors can significantly change the velocity distribution of the clusters in
the generated beam and therefore new systematic studies of these cluster target
beams had to be performed.

The mean velocity of the hydrogen clusters produced with a similar trumped shaped
nozzle with a throat diameter of $37\,\mu\text{m}$ was already measured\cite{Allspach1998} at
the target for the E835 experiment at FERMILAB
and it was found, that it can be described adequately by the maximum local velocity
$u_\text{max}$ of a perfect gas accelerated in an isentropic expansion
through a convergent-divergent nozzle\cite{Allspach1998,Christen2010}:
  \begin{equation}
    u_\text{max} = \sqrt{\frac{2\,\kappa}{\kappa-1}\frac{R\,T_0}{M}}\,.
		\label{eq:MaximumVelocityPerfectGas}
  \end{equation}
In this equation $T_0$ is the gas temperature at the inlet of the nozzle and
$\kappa$ is the adiabatic index. These measurements were done at a pressure
below $8\,\text{bar}$ and temperatures between $15\,\text{K}$ and
$40\,\text{K}$ where the hydrogen is in the gas phase before entering the nozzle.
However, later optimization studies on hydrogen cluster targets\cite{Taeschner2007,Taeschner2011} 
showed that a significant performance increase is possible, i.e., an increase of the achievable
maximum target thickness by orders of magnitude. One prerequisite for
this is that the target is operated with hydrogen being in the liquid phase before 
entering the nozzle.
Since it is
known from previous measurements, e.g., Ref.~\onlinecite{Knuth1995}, that the
phase transition from the gas phase into the liquid phase has a significant impact on the cluster velocity,
precise measurements on the velocity distributions and mean velocities as
well as a comparison with the situation obtained with hydrogen being in
the gas phase before entering the nozzle were strongly needed. Based on this
data verifications and optimizations of calculations will be possible.

For this reason a dedicated time-of-flight system was designed and installed at the 
M\"{u}nster cluster jet target. Detailed studies on the velocity distributions
as function of the operating parameters were performed which is presented in the
first part of this work.

Although the time-of-flight system enabled for a precise determination of the
velocity distributions and with this for the calculation of the mean velocities,
the measurement time of several hours makes it impractical to use this
system regularly for the volume density determination described above. This is especially
true for optimization studies where the temperature and pressure at the nozzle
inlet is changed very often. Therefore it was essential to be able to calculate
the mean cluster velocity as function of the stagnation conditions in the
typical operation region with a precision, as discussed before, below 10\%.
Many groups, e.g.,  Ref.~\onlinecite{Knuth1995},
Ref.~\onlinecite{Buchenau1990}, and Ref.~\onlinecite{Christen2013} have 
measured mean velocities of clusters and extracted fluid properties like
the temperature of the cluster at the end of the expansion based on different
equations of state. Ref.~\onlinecite{Harms1997}, for example, used an equation of
state to produce a theoretical prediction for the mean velocity by assuming a
constant final temperature. The values predicted with this method deviate
from the measurements presented in the same work by about 20\%--40\%
for the data points measured at temperatures below the boiling point. 
Therefore a new method had to be developed to allow for more accurate predictions
at the discussed stagnation conditions.
In the second part of this work such a method is presented, introducing two
parameters which had to be determined only once by a fit to the measured data.
In the operation region of the investigated cluster source this techniques
provides precise predictions with an average absolute deviation
of only about 5\% compared to the measured mean cluster velocities.
This applies both to the regions of liquid and of gaseous hydrogen in front of
the nozzle.

\section{Experimental setup}

\begin{figure}[b]
\includegraphics[width=\columnwidth]{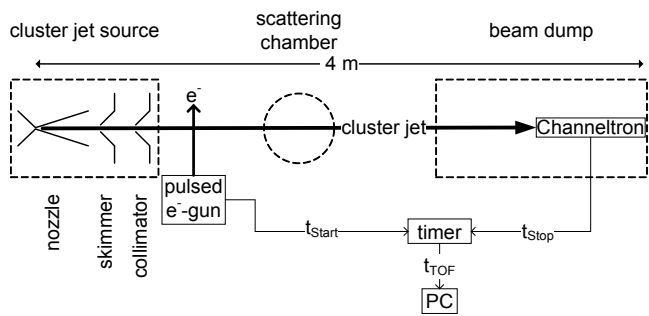}
\caption{Schematic view of the time-of-flight setup for the velocity
 determination of single clusters.}
\label{fig:TOFSetup}
\end{figure}
In Fig.~\ref{fig:TOFSetup} the schematic view of the utilized time-of-flight
setup is shown. Clusters produced in the cluster source are ionized by a
pulsed electron gun mounted at a short distance behind the collimator. This
electron gun is operated in a pulsed mode with a repetition rate of 
about $20\,\text{Hz}$ and a pulse width of approximately $20\,\mu\text{s}$. 
The current of the electron beam is reduced in such a way that for each pulse no
more than a single ionized cluster is registered. The ionized clusters itself are 
detected by a Channeltron after a flight path of $4.07\pm0.02\,\text{m}$.
Due to this long distance 
and pulse widths in the microsecond time scale, the observed time-of-flight times 
being in the range between $4\,\text{ms}$ and $26\,\text{ms}$ 
can be obtained with high resolution.
The start and stop pulses are detected by a timer system
based on a MC9S08QG8 micro controller by Freescale Semiconductor and the time
difference is send to a computer. A detailed description of the used software
for the micro controller can be found in Ref.~\onlinecite{Taeschner2013}.

\begin{figure}
\includegraphics[width=\columnwidth]{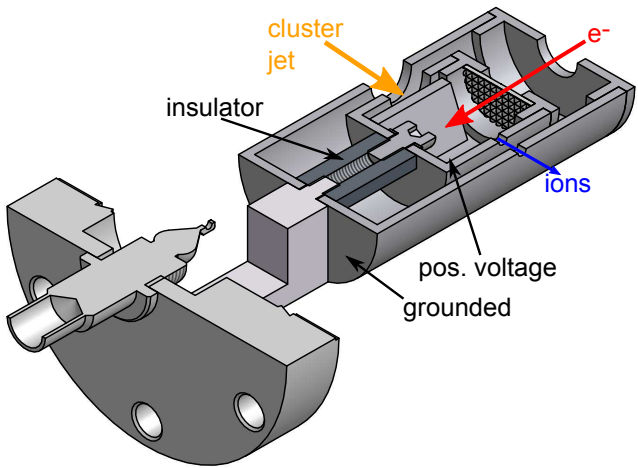}
\caption{\label{fig:CalibrationSource}Cross section of the
calibration source used to produce hydrogen ions with known kinetic energies.}
\end{figure}
In order to extract time-of-flight information with high resolution  
the complete setup has to be calibrated with respect to possible timing
offsets introduced by the pulsed electron gun device and the cluster detection
system. 
For this purpose the calibration source for ions with known kinetic energy is used (Fig.~\ref{fig:CalibrationSource}).
The source consists of two coaxial cylinders. The outer cylinder is electrically
grounded while the inner cylinder is connected to a voltage source providing a potential of up to
$4\,\text{kV}$. The calibration source is mounted in such a way,
that the cylinder axis is perpendicular to the axis of the incoming cluster beam,
so that the clusters can enter through a hole with a diameter of
$10\,\text{mm}$. In the inner cylinder the clusters are stopped by a wedge
shaped plate, evaporate and are converted into hydrogen gas. This gas is ionized by the pulsed electron
beam entering along the axis of the two cylinders. The potential difference
between the cylinders accelerate the produced ions to a known kinetic energy 
while they are extracted through a $2\,\text{mm}$ hole along the cluster beam axis.
The ions leaving this unit drift towards the detection system which consists of an array
of a grounded entrance orifice followed by a Channeltron. Here the Channeltron input 
is set on a negative potential of $2.1\,\text{kV}$ relative to the entrance orifice. 
Thus, positively charged ions and, in the later cluster
time-of-flight measurements, positively charged clusters are accelerated 
and cause detectable signals. 
 
\begin{figure}[b]
\includegraphics[width=\columnwidth]{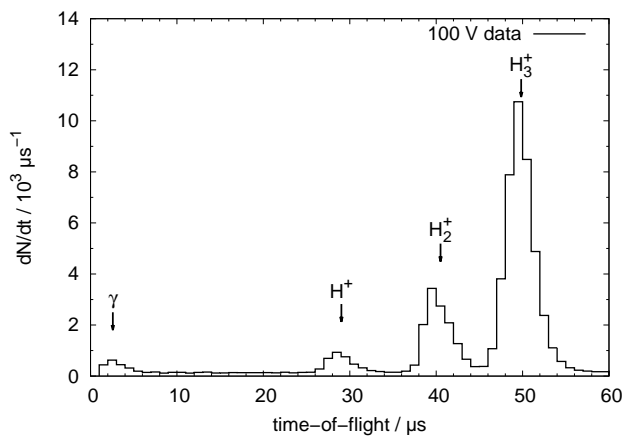}
\caption{\label{fig:TOFCalibrationDistribution}Example of a time-of-flight
distribution using the calibration source.}
\end{figure}
In Fig.~\ref{fig:TOFCalibrationDistribution} an example of a time-of-flight
distribution is shown before timing calibration which was measured using 
the calibration source with an
acceleration voltage of $100\,\text{V}$. In this distribution four peaks from
different ion species are clearly visible. The peak with the lowest mean
time-of-flight can be attributed to photons produced in the cluster source.
Since their time-of-flight is approximately $10\,\text{ns}$ the measured flight time
of $\sim 4\,\mu\text{s}$ is a direct measure of the timing offset
caused by the electronics. The three other peaks correspond to different
hydrogen ions, namely $\text{H}^+$, $\text{H}_2^+$, and $\text{H}_3^+$. The time offset and
the length of the flight path between the electron gun and the Channeltron could
be extracted by measuring the mean time-of-flight for the different ions
as function of the acceleration voltage. For these measurements the
electron gun was operated at a repetition rate of about $25\,\text{kHz}$ and
a pulse duration of about $2\,\mu\text{s}$. With this calibration setup
a time resolution of about $3\,\mu\text{s}$ was reached which is 
predominantly given by the pulse duration of the electron gun.

\section{Velocity distributions of clusters}

Using the presented time-of-flight setup the velocity distributions of hydrogen
clusters produced in the M\"{u}nster cluster jet target setup
were measured\cite{Koehler2010} using a nozzle with a minimum diameter of 
$28\,\mu\text{m}$. For these measurements the pulse duration of the electron gun
was increased to $20\,\mu\text{s}$, so that the time resolution increases to about
$21\,\mu\text{s}$ which is still very precise compared to the measured 
standard deviation of the typical velocity destributions of the clusters of
several hundred to thousand microseconds.

\begin{figure}[b]
  \includegraphics[width=\columnwidth]{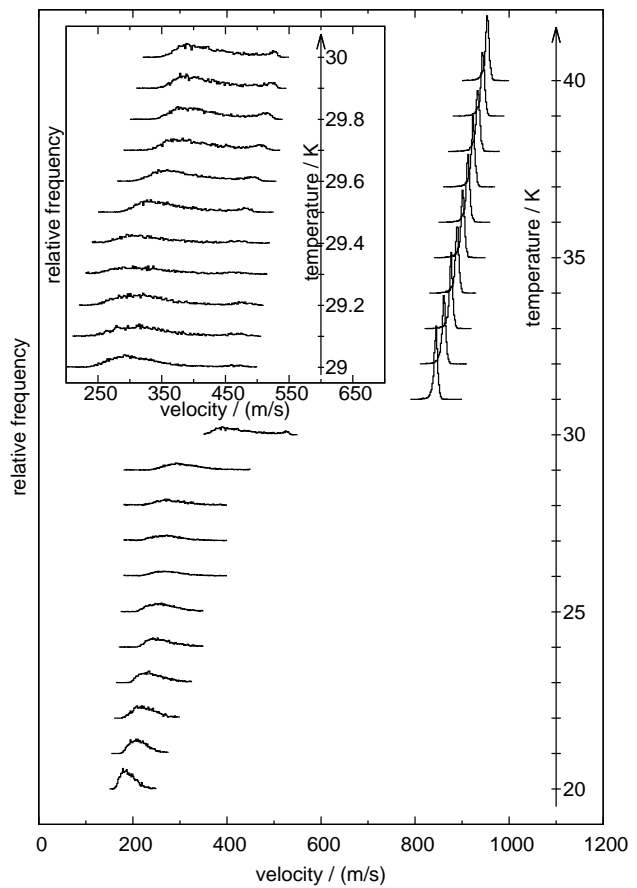}
  \caption{\label{fig:VelocityDistrib8bar}Distributions of the cluster velocities
	as function of the inlet temperature of the fluid at a constant pressure
	before the nozzle of $8\,\text{bar}$. The inlayed graph shows a zoom
	into a small temperature range.}
\end{figure}
\begin{figure}
  \includegraphics[width=\columnwidth]{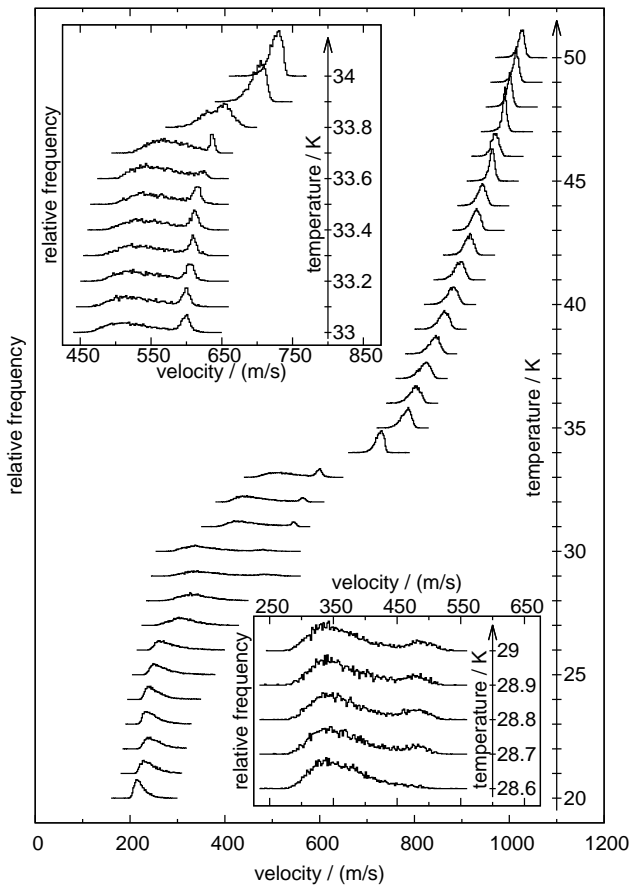}
  \caption{\label{fig:VelocityDistrib14bar}Distributions of the cluster velocities
	as function of the inlet temperature of the fluid at a constant pressure
	before the nozzle of $14\,\text{bar}$. The inlayed graphs show a zoom
	into two small temperature ranges.}
\end{figure}
In Fig.~\ref{fig:VelocityDistrib8bar} and \ref{fig:VelocityDistrib14bar} the
measured distributions of the cluster velocity are shown. In the first figure
a constant pressure of $8\,\text{bar}$ was used in front of the nozzle and for the
second figure $14\,\text{bar}$ were applied. In both figures the distributions at
different fluid temperatures from $20\,\text{K}$ up to $50\,\text{K}$ are
displayed. The distributions are scaled in such a way that the total area of
each spectrum is the same in the respective figure.

In case of Fig.~\ref{fig:VelocityDistrib14bar} where the data at $14\,\text{bar}$
are shown the distributions above the boiling point of around $34\,\text{K}$ are
relatively sharp with a standard deviation of about $10\,\text{m/s}$ and have
a negative skew. At the phase transition between gas and liquid a double peak
structure is visible with a narrow peak at higher mean velocity on top of
a broad peak with a lower mean velocity. The inlayed graph shows the development of
this structure in the temperature range between $33\,\text{K}$ and $34\,\text{K}$.
At a temperature of around $33.7\,\text{K}$ the width of the smaller peak is
around $3\,\text{m/s}$ at a mean velocity of about $637\,\text{m/s}$ whereas
the broader peak has a standard deviation of about $25\,\text{m/s}$ and a
mean velocity of $576\,\text{m/s}$.

This double peak structure was also observed in earlier experiments of other groups,
presented, e.g., in Ref.~\onlinecite{Knuth1995}, \onlinecite{Buchenau1990},
and \onlinecite{Harms1997}, both with hydrogen and helium.
The cited authors explain their observations with a different kind of
cluster production for the respective peak. They expect that the narrow peak
consists of clusters which were formed from a gas by condensation
whereas the broad peak should consist of clusters formed from a liquid by fragmentation.

Below $29\,\text{K}$ the double peak structure disappears which can be seen
in the second inlayed graph showing the measured distribution in the temperature
range between $28.6\,\text{K}$ and $33\,\text{K}$. At a temperature of
$28.7\,\text{K}$ the width of the narrow peak has increased to about
$17\,\text{m/s}$ in combination with a further reduced mean velocity of $478\,\text{m/s}$,
while the standard deviation of the broader peak has increased to $43\,\text{m/s}$
at a mean velocity of about $363\,\text{m/s}$. At lower temperatures only the
broader peak remains although the width of the peak decreases with a further
reduction of the fluid temperature from $43\,\text{m/s}$ at $28.6\,\text{K}$
to $13\,\text{m/s}$ at $20\,\text{K}$. The described behavior is similar at
different other pressures in front of the nozzle as can be seen in
Fig.~\ref{fig:VelocityDistrib8bar}.

\section{Mean cluster velocities}

As mentioned before, the precise knowledge of the mean cluster velocity $u$
is needed for the estimation of the volume density of the cluster beam. Since
it is not feasible to measure this quantity for each possible stagnation
condition within the operation region of the cluster target, a method is needed
to predict these values with an accuracy below about 10\%.
In contrast to other publications, e.g., Ref.~\onlinecite{Knuth1995}, the main
focus lies here on the description of the mean cluster velocity of the complete
velocity distribution which in our case can be directly calculated from the
measured velocities of the single clusters. We therefore will not quote the
mean velocity of the two peaks observed in the phase transition region separately.
In our case, where these two peaks completely overlap there is also no model
independent way to extract this information.

\subsection{Experimental results}

\begin{figure}[b]
\includegraphics[width=\columnwidth]{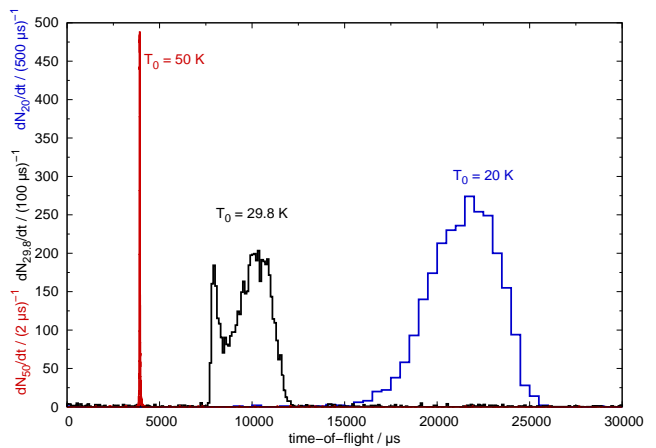}
\caption{\label{fig:TOFDistributionExamples}Distribution of the time-of-flight of hydrogen clusters
produced at three different hydrogen temperatures at the nozzle inlet and
with the same inlet pressure of $8\,\text{bar}$.}
\end{figure}
\begin{figure}
\includegraphics[width=\columnwidth]{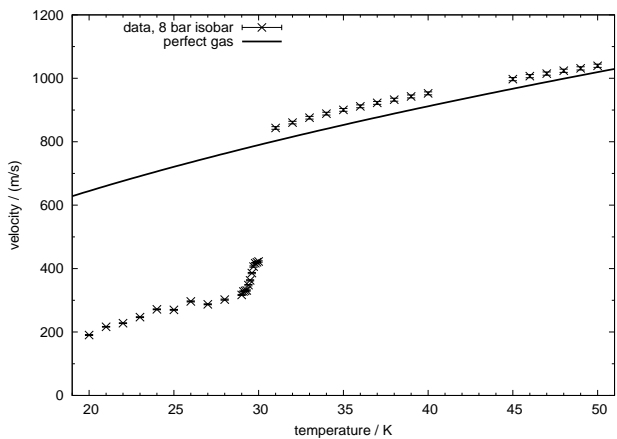}
\caption{\label{fig:MeanVelocitiesData}Mean velocity of hydrogen clusters as
function of the temperature at the nozzle inlet and with the same inlet pressure
of $8\,\text{bar}$. The solid line shows the maximum gas velocity of
perfect gas according to Eq.~(\ref{eq:MaximumVelocityPerfectGas}).}
\end{figure}

To summarize the above findings three examples for the measured time-of-flight
distributions of hydrogen clusters are presented in Fig.~\ref{fig:TOFDistributionExamples}.
The distributions were measured at a constant hydrogen pressure of $8\,\text{bar}$
at the nozzle inlet, but at different temperatures of $20\,\text{K}$,
$29.8\,\text{K}$, and $50\,\text{K}$. The distribution at highest temperature
is very narrow with a standard deviation of $\sim 19\,\mu\text{s}$ which
increases up to $1800\,\mu\text{s}$ at $20\,\text{K}$. Near the boiling
temperature of $30\,\text{K}$ a double peak structure was observed with a small peak
at higher velocity on top of a broad peak with a lower mean velocity indicating
the different cluster production mechanism from the two coexisting phases.
This dependance of the production process on the hydrogen phase state is also
visible in Fig.~\ref{fig:MeanVelocitiesData} where the mean cluster velocity
is plotted as function of the hydrogen temperature at the inlet of the nozzle
at a constant inlet pressure of $8\,\text{bar}$.

Above the boiling point the data can be described adequately by the maximum gas
velocity of a perfect gas (Eq.~(\ref{eq:MaximumVelocityPerfectGas})), which
agrees well with the observations presented in Ref.~\onlinecite{Allspach1998}
taken at lower inlet pressure of below $5\,\text{bar}$. However, the data below
the boiling point deviate from the calculated ones by up to a factor of three,
which is in agreement with the results presented in Ref.~\onlinecite{Knuth1995}.

\begin{figure}
  \includegraphics[width=\columnwidth]{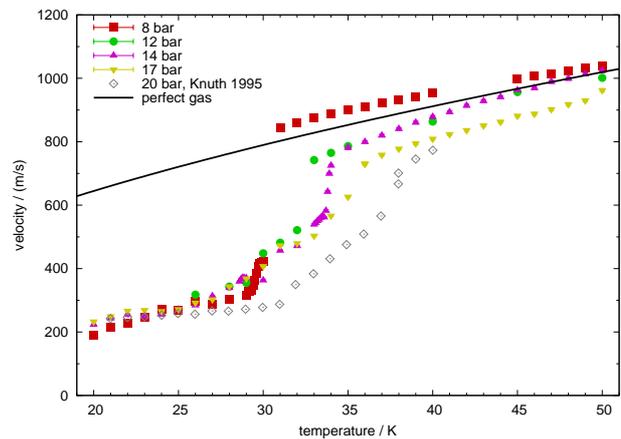}
  \caption{\label{fig:PerfectGasModelComparison}Measured mean hydrogen cluster
	velocities as function of the stagnation temperature for different
	constant stagnation pressures. For comparison the measurement presented
	in Ref.~\onlinecite{Knuth1995} obtained with a pinhole nozzle and a
	stagnation pressure of $20\,\text{bar}$ is shown. The solid line is
	calculated from Eq.~(\ref{eq:MaximumVelocityPerfectGas}) assuming a perfect gas.}
\end{figure}

A collection of measured mean velocities for different isobars is
presented in Fig.~\ref{fig:PerfectGasModelComparison}. Since the boiling point
shifts towards higher temperatures for higher stagnation pressures the
transition between high velocity to low velocity shifts accordingly.
For comparison, in this graph the data presented in Ref.~\onlinecite{Knuth1995}
is also displayed. In the temperature region where two peaks are observed,
only the dominating peak (Peak 4 in Ref.~\onlinecite{Knuth1995}) is shown for
better comparison, since we discuss here only the mean velocity of the clusters.
As mentioned before, our data are in good agreement with the results of
Ref.~\onlinecite{Knuth1995} although a pin hole
nozzle with a minimum diameter of $5\,\mu\text{m}$ was used there. A more
detailed discussion of this data is given in Sec.~\ref{sec:ModelCalculations}.

In Ref.~\onlinecite{Harms1997} it is shown that the cluster
velocity can be predicted by calculating the fluid velocity based on isentropic
expansion from the stagnation temperature down to the temperature of the
triple point. However, the calculations presented there deviate from the measured
data in the region below the boiling point by about 20\%--40\%,
which is too large for the application discussed here. The data
presented in Ref.~\onlinecite{Harms1997} and~\onlinecite{Knuth1995}
indicate that the temperature at the end of the expansion, the so called terminal
temperature, is dependent on the stagnation conditions and is found to be always
above the triple point temperature. Based on this knowledge and the fact that the used nozzle
in our case is comparably long, one can expect that the terminal temperature
is already reached inside the nozzle. Therefore, it is essential to calculate
the velocity of the fluid inside the nozzle as function of the distance from
the nozzle throat. As will be discussed below it is possible to introduce
two new parameters for these calculations which can be fixed by only two
velocity measurements. This enables the precise prediction of the cluster
velocities $u(p_{0}, T_{0})$.

Since above the boiling point the data are already described well using the
simple assumption of a perfect gas, the method for calculating the local velocity
inside the nozzle is explained first using this simple model.
In a next step the calculations will be done with an equation
of state which can describe a fluid with a gaseous and a liquid phase. 

\subsection{Model calculations}\label{sec:ModelCalculations}
\begin{figure}
  \centering
  \includegraphics[width=0.8\columnwidth]{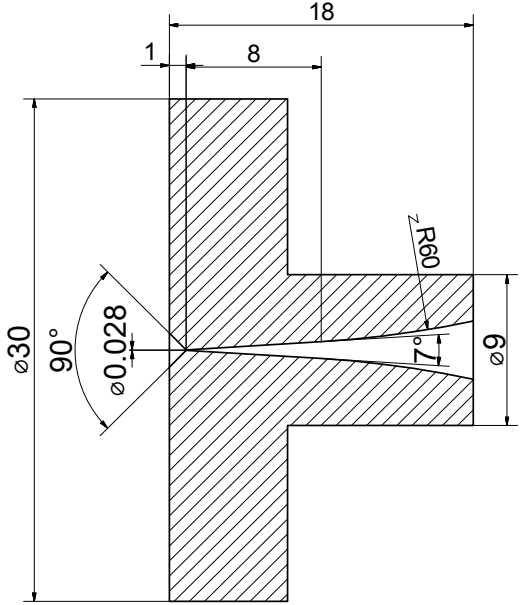}
  \caption{\label{fig:Nozzle}Cross section of the used Laval nozzle
	manufactured in the CERN workshop from copper.}
\end{figure}
In order to calculate the position dependent properties of the hydrogen 
fluid inside the nozzle, a stationary quasi-one-dimensional model is used.
The details of these calculations, which are based on the dimensions
of the used Laval nozzle shown in Fig.~\ref{fig:Nozzle}, are presented in
Appendix~\ref{app:LocalProperties}.
	
\begin{figure}[b]
    \includegraphics[width=\columnwidth]{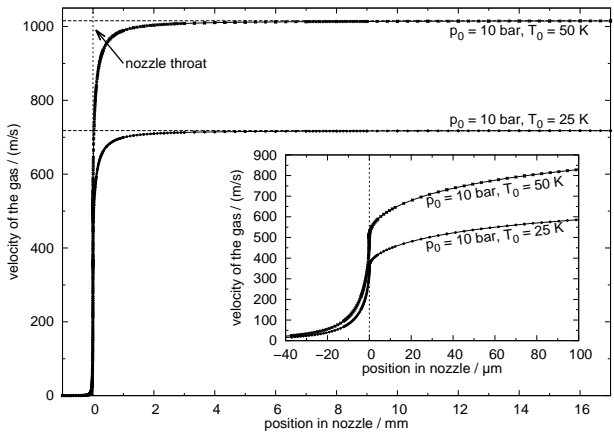}
    \caption{\label{fig:LocalVelocityPerfectGas}Local velocity of the perfect gas as
		function of the position inside the CERN nozzle for two different stagnation
		conditions. The two dashed horizontal lines correspond to the maximum gas
		velocity.}
\end{figure}

The applied method can be used with any equation of state. The simplest model
is the perfect gas with the following equation of state\cite{Anderson1990}:
\begin{equation}
  p = \rho\,R_{\text{s}}\,T\, ,
	\label{eq:EOSPerfectGas}
\end{equation}
where $p$ is the pressure and $R_{\text{s}}=R/M$ is the specific gas constant. In
contrast to the ideal gas, which has the same equation of state, the specific
heat at constant pressure or volume is constant in the case of the perfect gas.
In Fig.~\ref{fig:LocalVelocityPerfectGas} the calculated local velocity is shown 
for a perfect gas as
function of the position inside the nozzle. The two curves correspond to two
different stagnation temperatures at the nozzle inlets, namely, $25\,\text{K}$
and $50\,\text{K}$. For both curves a stagnation pressure of $10\,\text{bar}$
was assumed. It is obvious that a few millimeters behind the nozzle throat the
velocity is almost constant and at its maximum value. The limit of the local velocity 
$u_{\text{max}}$ is reached if the ratio $A(T_{z})/A^{*}$ between
the local area and the area of the throat approaches infinity and can be expressed by 
Eq.~(\ref{eq:MaximumVelocityPerfectGas}). The two horizontal dashed lines in
Fig.~\ref{fig:LocalVelocityPerfectGas} indicate the value of this maximum
velocity $u_{\text{max}}$ for the two stagnation temperatures.

\begin{figure}
\includegraphics[width=\columnwidth]{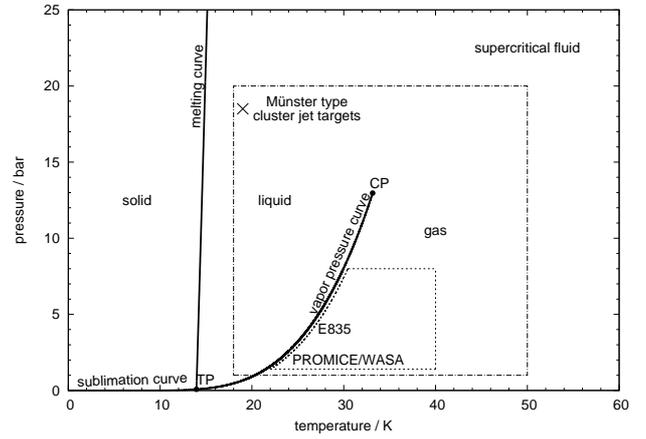}
\caption{Phase diagram of hydrogen with the triple point (TP),
  the critical point (CP), the vapor pressure curve, the melting curve, and
	the sublimation curve based on Ref.~\onlinecite{McCarty1981} and 
	Ref.~\onlinecite{Leachman2009}. The typical operation region of the conventional
	cluster jet targets of the PROMICE/WASA\cite{Ekstroem1995} and the
	E835\cite{Allspach1998} experiments are indicated by a dashed line
	and for the M\"{u}nster type target with a dashed dotted line. The point where
	the highest target thickness was obtained with the M\"{u}nster type
	target\cite{Taeschner2011} is indicated by a cross.}
\label{fig:PhaseDiagram}
\end{figure}
Since in case of the perfect gas the local velocity is already almost constant a few millimeters
behind the nozzle throat, the maximum velocity $u_{\text{max}}$ can be used as
a first order estimate for the mean cluster velocity. In
Fig.~\ref{fig:PerfectGasModelComparison} a corresponding model calculation is compared to
measurements with different stagnation pressures. Above a certain temperature,
which changes depending on the stagnation pressure, the model agrees well with
the measured data, but for lower temperatures the measured velocities are up to a
factor of three lower than the model predictions. 
Comparing these specific temperatures with the phase diagram of hydrogen shown in
Fig.~\ref{fig:PhaseDiagram} it can be seen, that for pressures below the
critical pressure these temperatures can be identified as the pressure
dependent boiling temperatures of normal hydrogen.

As mentioned above, the perfect gas equation of state was only used to
explain the method used for calculating the local velocity inside the nozzle.
The simplest model which describes
a fluid with both a gaseous and a liquid phase is the van der Waals gas with the
following equation of state:
\begin{equation}
  p = \frac{R_{\text{s}}\,T}{v-b^{\prime}} - \frac{a^{\prime}}{v^{2}},
	\label{eq:EOSVanDerWaals}
\end{equation}
with the specific volume $v=1/\rho$ and the two constants $a^{\prime}$ and
$b^{\prime}$ which can be calculated from the pressure and temperature at the
critical point of the used gas (see, e.g., Ref.~\onlinecite{Nolting4}).
In case of hydrogen a critical temperature of $T_{\text{c}}=33.19\,\text{K}$
and a critical pressure of $p_{\text{c}}=13.15\,\text{bar}$ were used,
which were taken from Ref.~\onlinecite{McCarty1981}. The detailed description
of the method used to calculate the local properties for the van der Waals gas
is given in Appendix~\ref{app:VanDerWaals}.

\begin{figure}
    \includegraphics[width=\columnwidth]{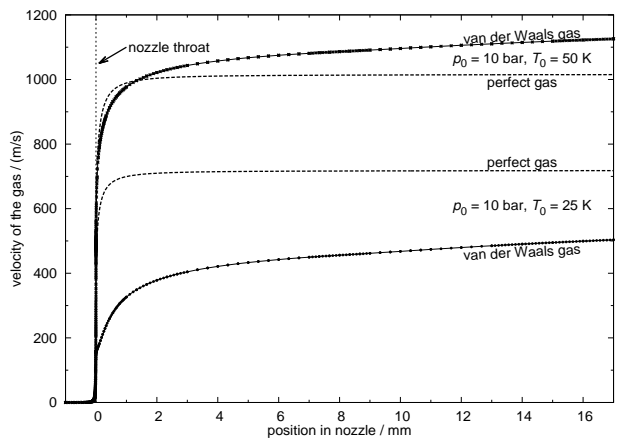}
    \caption{\label{fig:LocalVelocityVanDerWaals}Local velocity of the fluid
		based on the van der Waals model (solid lines) and the perfect gas (dashed line) 
		as function of the position inside the CERN nozzle for two different stagnation
		conditions.}
\end{figure}

In Fig.~\ref{fig:LocalVelocityVanDerWaals} the calculated local velocity
is presented as function of the position inside the nozzle for two different
combinations of stagnation pressure and temperature. The dashed lines
show the solutions for the perfect gas and the solid line the solutions for the
van der Waals equation of state. It is clearly visible that the value of the
local velocity does not saturate for the van der Waals model in contrast to
the values calculated for the perfect gas. Since it was observed by other
groups, e.g., Ref.~\onlinecite{Harms1997} and Ref.~\onlinecite{Knuth1995}, that
the temperature at the end of the expansion path of the cluster formation is
strongly dependent on the stagnation conditions, this terminal temperature cannot
be used to predict precisely the mean velocity of the clusters. Instead,
we chose here the position inside the nozzle as new parameter to produce such
a prediction. This choice can be motivated by the production process of clusters
which are formed by condensation from a gas. In this case it is obvious
that at a certain point inside the nozzle the mean number of collisions between
the clusters and the surrounding molecules is so low and the mass of the
clusters so high that further collisions do not change the mean velocity anymore.

\begin{figure}[b]
    \includegraphics[width=\columnwidth]{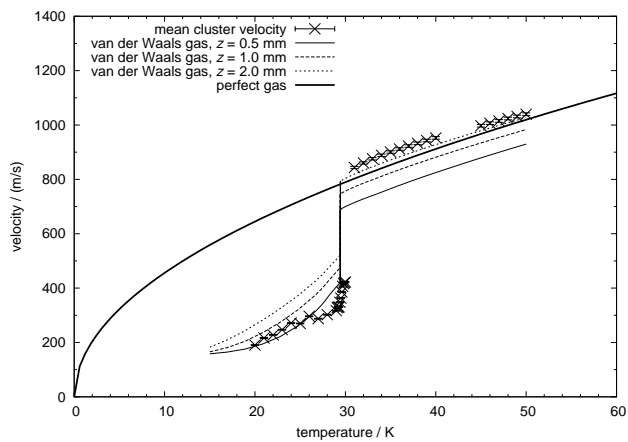}
    \caption{\label{fig:VelocityIsobar8bar}Mean cluster velocity as
	  function of the stagnation temperature for an isobar at 8 bar.
		The solid line is calculated assuming a perfect gas whereas the other
		lines represent the local velocity at three different positions of
    $0.5\,\text{mm}$, $1\,\text{mm}$, and $2\,\text{mm}$ behind the nozzle throat.}
\end{figure}
\begin{figure}
    \includegraphics[width=\columnwidth]{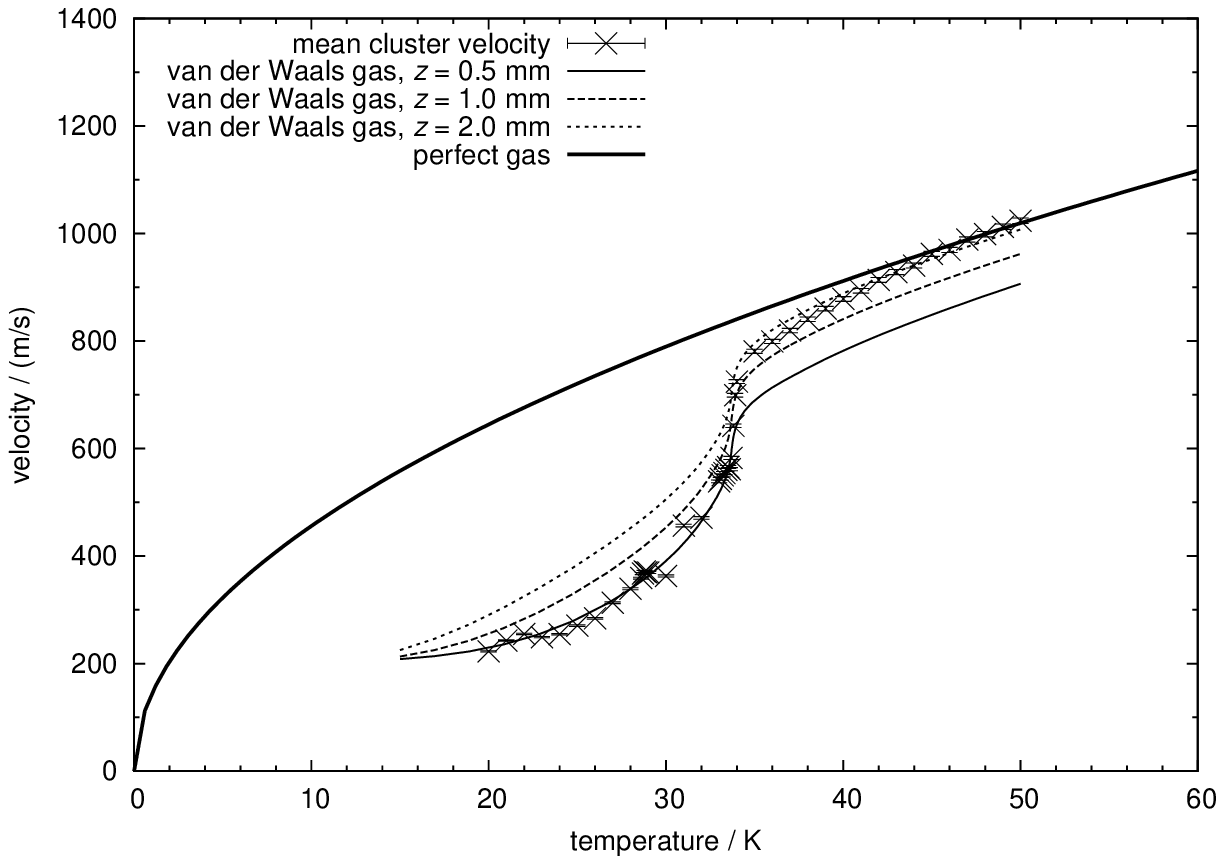}
    \caption{\label{fig:VelocityIsobar14bar}Mean cluster velocity as
	  function of the stagnation temperature for an isobar at 14~bar.
		The solid line is calculated assuming a perfect gas whereas the other
		lines represent the local velocity at three different positions of
    $0.5\,\text{mm}$, $1\,\text{mm}$, and $2\,\text{mm}$ behind the nozzle throat.}
\end{figure}
\begin{figure}[b]
    \includegraphics[width=\columnwidth]{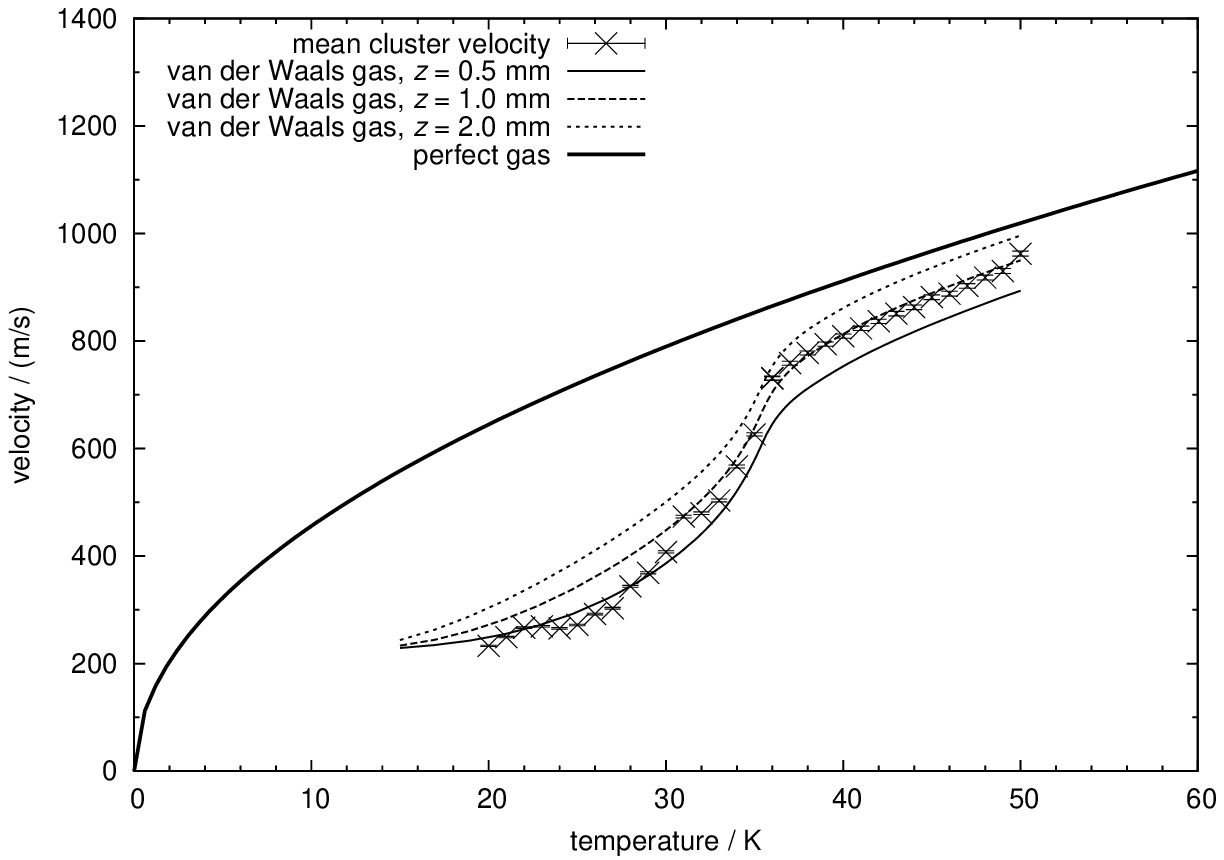}
    \caption{\label{fig:VelocityIsobar17bar}Mean cluster velocity as
	  function of the stagnation temperature for an isobar at 17~bar.
		The solid line is calculated assuming a perfect gas whereas the other
		lines represent the local velocity at three different positions of
    $0.5\,\text{mm}$, $1\,\text{mm}$, and $2\,\text{mm}$ behind the nozzle throat.}
\end{figure}
\begin{figure}
    \includegraphics[width=\columnwidth]{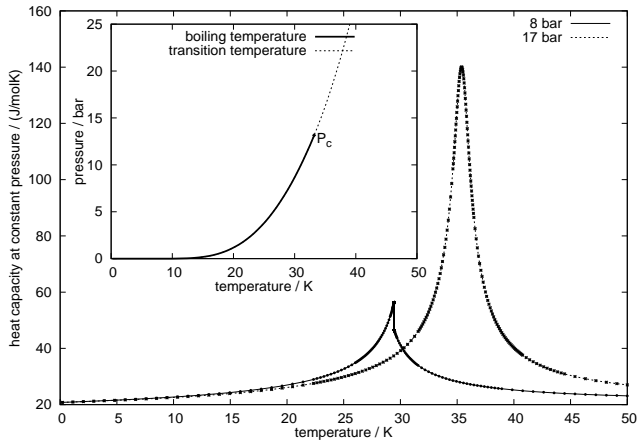}
    \caption{\label{fig:TransitionTemperature}Calculated heat capacity $c_p$
		as function of the temperature based on the van der Waals model
		for two different isobars at 8~bar and at 17~bar. In the inlay figure the
		transition temperature defined as the position of the maximal heat capacity is
		displayed together with the vapor pressure curve.}
\end{figure}
In Fig.~\ref{fig:VelocityIsobar8bar}--\ref{fig:VelocityIsobar17bar} the measured mean cluster velocities for different 
isobars at $8\,\text{bar}$, $14\,\text{bar}$, and $17\,\text{bar}$ are compared
to calculated local velocities at three different positions of
$0.5\,\text{mm}$, $1\,\text{mm}$, and $2\,\text{mm}$ behind the nozzle throat.
It is obvious that the measured data
can be described in good approximation by the calculated velocities if a
position of $0.5\,\text{mm}$ is used for the data taken at temperatures
below the boiling point and of about $1.5\,\text{mm}$ above this point.

Based on this observation two parameters, i.e.,
${z_{\text{l}}\approx0.5\,\text{mm}}$ and $z_{\text{g}}\approx1.5\,\text{mm}$,
were introduced,
specifying the position inside the nozzle where the local velocity
$u_{\text{vdW}}(p_{0}, T_{0}, z)$ is calculated. The mean cluster velocity
can then be predicted using the following equation:
\begin{equation}
  u_{\text{C}} = \begin{cases}
    u_{\text{vdW}}(p_{0}, T_{0}, z_{\text{l}}) &
      \text{for } T_{0} < T_{\text{tr}}(p_{0}) \\
    u_{\text{vdW}}(p_{0}, T_{0}, z_{\text{g}}) &
      \text{for } T_{0} \geq T_{\text{tr}}(p_{0})
  \end{cases}\, .
  \label{eq:ClusteVelocityFit}
\end{equation}
The cluster production mechanism differs depending on the phase state
of the fluid in front of the nozzle. In case of a gas at the inlet the clusters
are formed by condensation and in case of a liquid from breakup and evaporation. Therefore,
it is plausible that for the two mechanisms, having a completely different
expansion path as indicated in Fig.~\ref{fig:DensityVanDerWaals}, one has to
allow for two different values for the position $z$. This is done in
the above equation by the two parameters $z_{\text{l}}$ and $z_{\text{g}}$.
For pressures $p_{0}$ below the critical pressure of $p_{\text{c}}=13.15\,\text{bar}$,
the specific temperature $T_{\text{tr}}(p_{0})$, which is used to switch between the
two velocity regimes, is the pressure dependent boiling temperature.
For pressures above this point the phase transition between gas and liquid is
continuous, so that there is no explicit boiling temperature. Nevertheless,
for the method described here such a well defined temperature is needed. Obviously
there are different choices possible, however, an approach described in
Ref.~\onlinecite{Sedunov2011} is well suited since the
transition temperatures produced by this method, displayed in the inlay of 
Fig.~\ref{fig:TransitionTemperature}, are an direct extension of
the vapor pressure curve to higher pressures. This transition temperature is defined as the
temperature with the maximal heat capacity at constant pressure
$c_p=(\partial h/\partial T)_p$.
In Fig.~\ref{fig:TransitionTemperature} the temperature dependence of this
heat capacity is displayed for one isobar below and for one
above the critical pressure. For pressures below the critical pressure
the boiling temperature is directly visible as a discontinuity of the
heat capacity, whereas in case of pressures above the critical pressure the
heat capacity exhibits a clear maximum. The position parameters $z_{\text{l}}$
and $z_{\text{g}}$ are adjusted in such a way that the deviation between the
predicted velocity $u_{\text{C}}$ and the measured mean cluster velocities are
minimized. In case of the studied cluster jet target the best fit values of these
position parameters were ${z_{\text{l}}=0.445\pm0.014\,\text{mm}}$ and
${z_{\text{g}}=1.67\pm0.20\,\text{mm}}$. With these values the described method
produces very precise predictions for the observed mean cluster velocities with mean
absolute deviation of about 5.1\% between measured and predicted velocities.
Since the precision of this prediction is better than the above mentioned required
precision of 10\%, the use of more sophisticated equations of state,
which were used in the work of other groups, e.g., Ref.~\onlinecite{Christen2010},
was not required. Furthermore, the excellent agreement between the measured data and
the predictions suggest that the influence of the transition to the clustered phase
on the used thermodynamic parameters, e.g., entropy, are fully represented by
the choice of the two position parameters.

\begin{figure}
  \includegraphics[width=\columnwidth]{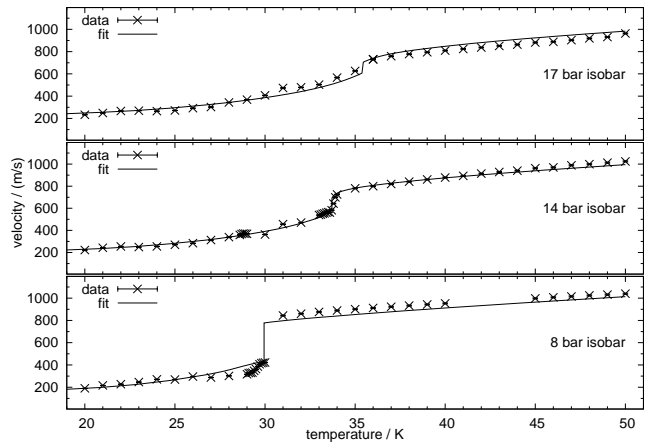}
  \caption{\label{fig:FittedVelocityIsobar}Comparison between the measured
	  mean cluster velocities at different isobars with the prediction made
		using Eq.~(\ref{eq:ClusteVelocityFit}).}
\end{figure}
\begin{figure}[b]
  \includegraphics[width=\columnwidth]{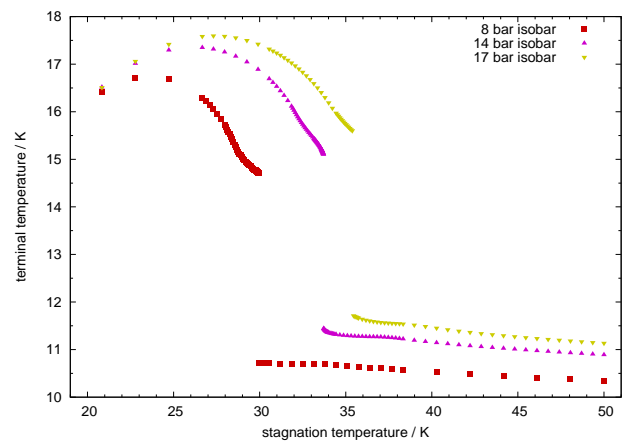}
  \caption{\label{fig:TerminalTemperature}Local temperature at different isobars
	  calculated at the same positions as the local velocities shown in
		Fig.~\ref{fig:FittedVelocityIsobar}.}
\end{figure}
\begin{figure}
  \includegraphics[width=\columnwidth]{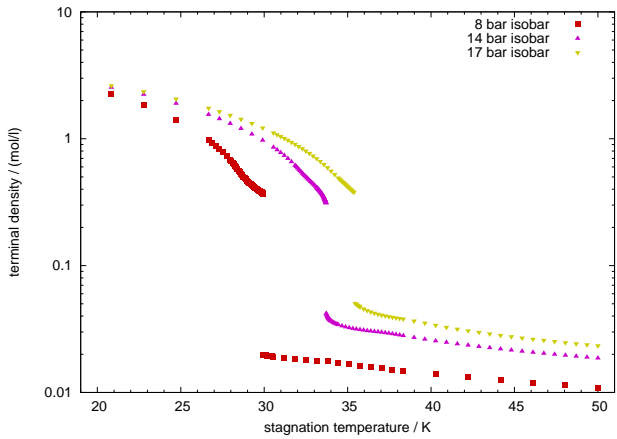}
  \caption{\label{fig:TerminalDensity}Local density at different isobars
	  calculated at the same positions as the local velocities shown in
		Fig.~\ref{fig:FittedVelocityIsobar}.}
\end{figure}
In Fig.~\ref{fig:FittedVelocityIsobar} the measured data for different isobars
are compared to the values calculated using Eq.~(\ref{eq:ClusteVelocityFit})
showing the good agreement for the different data sets.
As mentioned above the terminal temperature or density cannot be used
as a parameter for reaching the required precision. In Fig.~\ref{fig:TerminalTemperature}
the local temperatures inside the nozzle, calculated at the same positions 
as used for the local velocities shown in Fig.~\ref{fig:FittedVelocityIsobar},
are presented. It is obvious that these terminal temperatures are not constant
and, therefore, cannot be used as parameters for a velocity prediction here. The
values for the terminal temperature range between $14\,\text{K}$ and $18\,\text{K}$
for stagnation temperatures below the transition temperature and around
$11\,\text{K}$ above this temperature. This is in good agreement with the
values presented in Ref.~\onlinecite{Knuth1995}.
In Fig.~\ref{fig:TerminalDensity} the terminal densities are shown which
were calculated in the same way as the terminal temperatures. It is obvious
that also this parameter cannot be used to make sufficiently precise predictions
since it changes over an order of magnitude in the stagnation temperature region 
below the transition temperature. In summary, with the proposed position
parameters $z_{\text{l}}$ and $z_{\text{g}}$ a high predictive power
of the model calculation is reached. In order to further investigate this
observation detailed studies with different nozzle geometries are planned
in the future.

\begin{figure}[b]
  \includegraphics[width=\columnwidth]{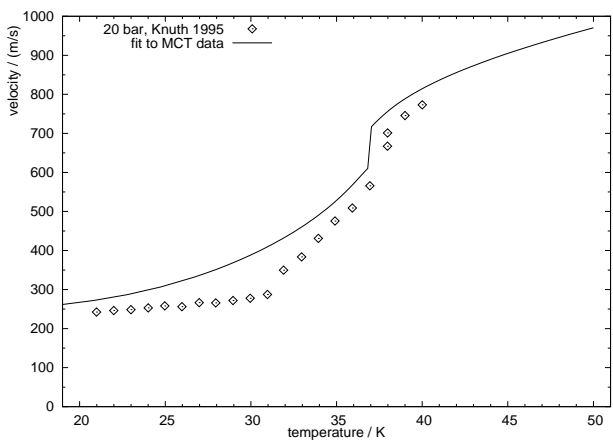}
  \caption{\label{fig:FittedVelocityKnuth}Comparison of the data presented
	in Ref.~\onlinecite{Knuth1995}, obtained with a pinhole nozzle and a
	stagnation pressure of $20\,\text{bar}$, and the prediction based on
	Eq.~(\ref{eq:ClusteVelocityFit}) using the fit parameters obtained from
	the data measured with the Laval nozzle presented in Fig.~\ref{fig:Nozzle}.}
\end{figure}
The relevance of the nozzle geometry itself might be illustrated by the
comparison of the presented model calculations with results from
Ref.~\onlinecite{Knuth1995} obtained with a pinhole nozzle with a minimum
diameter of $5\,\mu\text{m}$. Since the exact geometry of this pinhole
nozzle is not known, in Fig.~\ref{fig:FittedVelocityKnuth} the measured data is
compared to the values calculated using Eq.~(\ref{eq:ClusteVelocityFit}) based on the
fitted values for the position parameters presented above. Obviously the velocities measured
using a pinhole nozzle differ significantly, i.e., up to 50\%, from the calculated
ones. This indicates the relevance of the nozzle geometry, e.g., the length and
the shape of the exit trumpet, on the mean cluster velocity.

\section{Volume flow through the nozzle}

\begin{figure}
\includegraphics[width=\columnwidth]{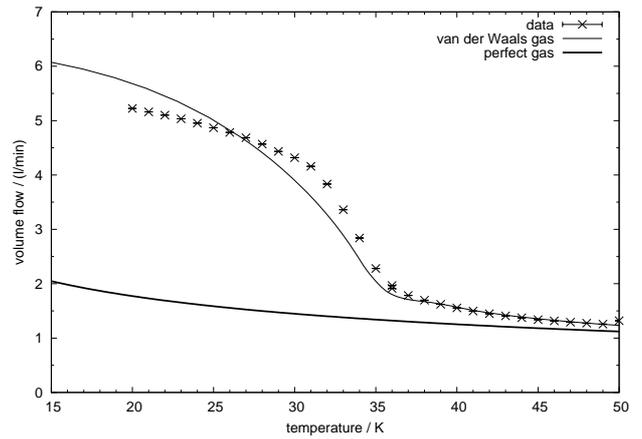}
\caption{\label{fig:VolumeFlow}Volume flow through the nozzle as function of
  the inlet temperature at a constant inlet pressure of $17\,\text{bar}$.}
\end{figure}
Using the method described above, not only the local properties but also global 
properties like the mass flow can be 
calculated from the critical properties using the following formula\cite{Anderson1990}:
  \begin{equation}
    \dot{m} = \rho^*\,u^*\,A^*\,.
  \end{equation}
In Fig.~\ref{fig:VolumeFlow} the measured volume flow towards the nozzle
is shown as function of the temperature in front of the nozzle using a
pressure of $17\,\text{bar}$. The volume flow was measured directly in
the gas supply line using a commercial mass flow meter. It is clearly
visible that the calculations based on the van der Waals model describe
the data very well especially above a temperature of around $38\,\text{K}$
whereas the calculations for the perfect gas fail to explain the data.
The largest deviations between the perfect gas model and the data are
visible below the transition temperature of about $35\,\text{K}$.
Contrary, a qualitatively good description is reached by the van der Waals although
it is also found that the data at lower temperatures are not fully described.

A similar discrepancy between calculations based on the van der Waals model and 
experimental data below the transition temperature was also observed by 
other groups which studied the flow of water vapor. A description is
given for example in
Ref.~\onlinecite{Sallet1991} where this observation was explained as a local deviation
from the thermal equilibrium caused by the finite evaporation rate of the liquid.
Models using rate equations can be used principally to calculate the flow in such a
case, however, the achieved precision of the discussed model is already 
sufficient for the desired investigations presented in this work.

\section{Summary}

For a precise measurement of the velocity of single hydrogen clusters produced
in the source of a high density cluster jet target
a time-of-flight setup using a pulsed electron gun was built up. A rich data sample for mean hydrogen cluster
velocities and velocity distributions are provided for different stagnation
conditions both above and below the critical pressure.

The mean values of the obtained cluster velocity distributions were compared
with model calculations based on both the equation of state of the perfect gas
and of the van der Waals gas. It was found that a precise prediction of the
measured data is possible by the van der Waals model if two cut-off position
parameters are introduced for which the local velocities inside the nozzle
are calculated. By adjusting these two parameters to the measured data, a
precise prediction of the mean cluster velocities is possible. In principle,
these two positions can be fixed by measuring one velocity at a temperature
above and one below the boiling point.
The average absolute deviation between the predicted velocities 
and the measured mean cluster velocities are found to be only about 5\%. 
Therefore, the approach presented in this work provides an excellent tool to 
predict the mean cluster velocities in a regime especially relevant for 
high intense cluster jet beams. For a specific nozzle, an essential
parameter required, e.g., for the determination of the absolute target thickness via
the scanning rod method, can be provided without further measurements.
In order to further investigate the observed excellent predictive power of the 
position parameters, a detailed study of the dependence of these
parameters on the nozzle geometry is planned.

\begin{acknowledgments}
The authors would like to thank H.~Orth for the very inspiring and
helpful discussions and H.~Baumeister and W.~Hassenmeier for their
support during the design of the target device. We are grateful to
M.~Macri and J.~Ritman for providing powerful vacuum pumps. The work
provided by the teams of our mechanical and electronic workshops is
very much appreciated and we thank them for the excellent
manufacturing of the various components. 
The research project was supported by BMBF (06MS253I and 06MS9149I/05P09MMFP8), GSI
F\&E program (MSKHOU1012), EU/FP6 HADRONPHYSICS (506078), EU/FP7
HADRONPHYSICS2 (227431), and EU/FP7 HADRONPHYSICS3 (283286).
\end{acknowledgments}

\appendix
\section{\label{app:LocalProperties}Calculation of local properties inside the nozzle}

In order to calculate the position dependent properties of the hydrogen 
fluid inside the nozzle, a stationary quasi-one-dimensional model is used.
Here the fluid properties are assumed to vary only along the $z$-axis, i.e.,
the symmetry axis of the nozzle, while the properties are considered to be constant 
in a plane perpendicular to the jet beam axis. 
This model implies that the flow is inviscid and without wall friction, external forces, and heat transfer
with the walls. Using these assumptions the flow has to be 
isentropic\cite{Anderson1990} and the following relation between the local cross section area~$A$
and velocity~$u$ can be derived\cite{Anderson1990}:
  \begin{equation}
    \frac{\text{d}A}{A} =
    \left(\mathit{M\!a}^{2}-1\right)\frac{\text{d}u}{u}\, ,
  \end{equation}
where $\mathit{M\!a}=u/a$ is the Mach number, which is the ratio between the
local velocity $u$ and the local speed of sound $a$. For this relation three different
cases can be discussed:

\begin{itemize}
	\item $\mathit{M\!a} < 1$: The flow is called subsonic. An increase of the
	local velocity ($du > 0$) is directly correlated with a decrease of the local
	area ($dA < 0$). A fluid flowing through a converging nozzle is therefore
	accelerated.
	\item $\mathit{M\!a} > 1$: The flow is called supersonic. An increase of the
	local velocity ($du > 0$) is directly correlated with an increase of the local
	area ($dA > 0$). A fluid flowing through a diverging nozzle is in this case
	also accelerated.
	\item $\mathit{M\!a} = 1$: The local velocity $u^{*}$ is equal to the local
	speed of sound $a^{*}$. In this case the local area is either at its maximum or
	minimum. For practical purposes only the case of minimal area is relevant.
\end{itemize}

Based on these considerations the flow through the used Laval nozzle 
is assumed to be subsonic in front of the nozzle throat
($z<0$) and supersonic behind it ($z>0$). At the position of the throat the Mach
number is exactly equal to one ($\mathit{M\!a}(z=0) = 1$). In the following text
all properties at the position, where the Mach number equals unity, the so called
critical properties, are marked with an asterisk (*). This assumption leads
directly to the knowledge of the size of the critical area $A^{*}$ which has to
be equal to the area of the nozzle throat $A_{t}=\pi\,r_{t}^{2}$, where
$r_{t}$ is the inner radius of the nozzle at its throat. In case of the
quasi-one-dimensional model the energy conservation can be expressed by the
following equation\cite{Anderson1990}:
\begin{equation}
  h_{1} + \frac{u_{1}^{2}}{2} = h_{2} + \frac{u_{2}^{2}}{2}\, ,
\end{equation}
where $h_{1,2}$ are the specific enthalpies at two positions inside the nozzle.
In case of the nozzle flow the velocity before the nozzle is considered to be
zero ($u_{0}=0$) so that the local velocity can be calculated from the following
formula:
\begin{equation}
  u(z) = \sqrt{2\,(h_{0}-h(z))}\, ,
  \label{eq:LocalVelocity}
\end{equation}
where $h_{0}=h(T_{0},\rho_{0})$ is the specific enthalpy before the nozzle and
$h(z)=h(T_{z},\rho_{z})$ is the local specific enthalpy at a position $z$
inside the nozzle. Here $T_{z}$ denotes the local temperature and $\rho_{z}$ the
local density at this position. The stagnation density $\rho_{0}$ can be
calculated from the stagnation pressure $p_{0}$ and the stagnation temperature
$T_{0}$ before the nozzle if the equation of state of the fluid is known.
Using this equation of state the specific enthalpy $h(T,\rho)$ and the
specific entropy $s(T,\rho)$ can be calculated. Since the flow is considered
to be isentropic the density~$\rho$ can be calculated directly from the
temperature~$T$ by solving the equation
\begin{equation}
  s(T,\rho) = s(T_{0},\rho_{0}).
\label{eq:SpecificEntropy}
\end{equation}
Therefore, Eq.~(\ref{eq:LocalVelocity}) is only dependent on the local
temperature $T_{z}$.

In order to calculate the local velocity the following steps have to be considered:
\begin{enumerate}
	\item Calculate the radius $r_{z} = r(z)$ of the nozzle at desired position
	$z$ and from this the area $A_{z}=\pi\,r_{z}^{2}$ at this position.
	\item Calculate the ratio $A_{z}/A_{t}$ of the local area $A_{z}$ and the
	area of the nozzle throat.
	\item Search for the temperature $T_{z}$ which satisfies the following equation:
	\begin{equation}
	  \frac{A_{z}}{A_{t}} = \frac{A(T_{z})}{A^{*}}\, .
	\label{eq:AreaRatio}
	\end{equation}
	This temperature is the local temperature at the desired position.
	\item Calculate the local velocity $u_{z}=u(T_{z})$.
\end{enumerate}

For the calculation of the local radius an analytic formula was derived from
the dimensions of the used Laval nozzle which is shown in Fig.~\ref{fig:Nozzle}.
The search for the temperature which satisfies Eq.~(\ref{eq:AreaRatio})
is most complex. It is done by a C\# port of the routine zeroin
\footnote{See http://www.netlib.org/go/zeroin.f for the original source code of this algorithm} which uses the
Brent Method\cite{NumericalRecipes2007} for the root finding. The limits of the temperature
interval are dependent on the flow type at the desired position. In front of the
nozzle throat the flow is subsonic, so that the temperature has to be between
the stagnation temperature $T_{0}$ in front of the nozzle and the critical
temperature $T^{*}$ $(T_{0}\geq T_{z}\geq T^{*})$. Behind the nozzle throat the
flow is supersonic and the temperature must be lower than the critical
temperature $T^{*}$ but larger than zero. In these numerical calculations a minimum temperature
$T_{\text{min}}>0$ has to be defined to guarantee that the density,
which is represented with floating point numbers with finite precision, is 
never equal to zero.

In case of the quasi-one-dimensional flow the mass conservation can be expressed
as continuity equation\cite{Anderson1990}:
\begin{equation}
  \rho_{1}\,u_{1}\,A_{1} = \rho_{2}\,u_{2}\,A_{2}\, ,
\end{equation}
where $\rho_{1,2}$ are the local densities at the two positions 1 and 2,
$u_{1,2}$ the local velocities and $A_{1,2}$ the local areas. From this
the ratio $A(T_{z})/A^{*}$ from Eq.~(\ref{eq:AreaRatio}) can directly be derived:
\begin{equation}
  \frac{A(T_{z})}{A^{*}} = \frac{\rho^{*}\,u^{*}}{\rho(T_{z})\,u(T_{z})}\, .
  \label{eq:AreaRatio2}
\end{equation}
In this equation $\rho^{*}$ is the critical density and $u^{*}$ the critical
velocity. The local density $\rho(T_{z})$ and velocity $u(T_{z})$ are calculated
by solving Eq.~(\ref{eq:SpecificEntropy}) and afterwards using
Eq.~(\ref{eq:LocalVelocity}).

\section{\label{app:VanDerWaals}Van der Waals equation of state}

\begin{figure}
\includegraphics[width=\columnwidth]{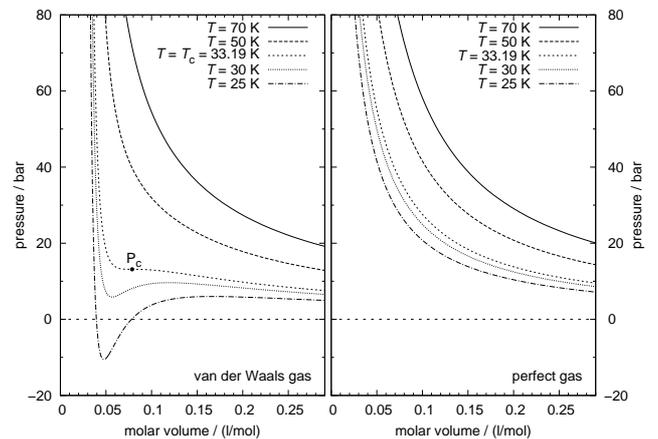}
\caption{Volume dependency of the pressure for selected isotherms for the
  van der Waals model of normal hydrogen (left) and the perfect gas (right).}
\label{fig:PressureEOSVanDerWaals}
\end{figure}

\begin{figure}[b]
\includegraphics[width=\columnwidth]{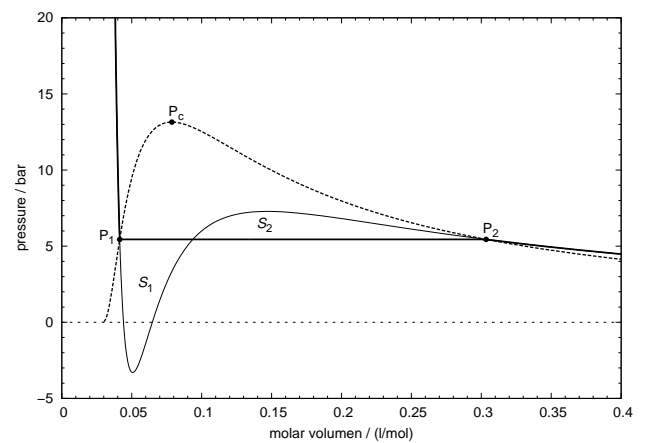}
\caption{Schematic usage of the Maxwell construction used to determine the
  vapor pressure in the coexistence region between the liquid and the gas phase.}
\label{fig:MaxwellConstruction}
\end{figure}

In case of the perfect gas equation of state (Eq.~(\ref{eq:EOSPerfectGas}))
a certain density or specific volume can be calculated
for each temperature and pressure value. This is not the case for the
van der Waals equation of state (Eq.~(\ref{eq:EOSVanDerWaals})).
Depending on the temperature up to three different values for the
density lead to the same pressure value. This can be seen in
Fig.~\ref{fig:PressureEOSVanDerWaals} where the pressure is shown as function of
the molar volume for different temperatures. Furthermore, it is obvious
that for certain temperatures this equation of state exhibits a behavior which
contradicts physics laws: it can lead to negative values for the pressure 
as well as for certain ranges of the molar volume where the pressure increases when the
volume is increased. Both can be cured using the well known Maxwell construction
shown in Fig.~\ref{fig:MaxwellConstruction}. Here the van der Waals
equation of state is replaced between the points $\text{P}_1$ and $\text{P}_2$
by a constant pressure, which is the vapor pressure at the given temperature.
There are multiple methods to find the temperature dependent points
$\text{P}_1$ and $\text{P}_2$. The most common one mentioned in text books
is to estimate a pressure between $\text{P}_1$ and $\text{P}_2$ that leads
to equally large areas $\text{S}_1$ and $\text{S}_2$. Although this method
is optimal to teach the concept of the Maxwell construction it has many
disadvantages when used in a numeric implementation. For this purpose
it is much more effective to use an equivalent method using the specific Gibbs
free energy $g$ and solving the following system of equations
(see, e.g., Ref.~\onlinecite{Nolting4}):
\begin{eqnarray}
  g(T,v_1) &=& g(T,v_2) \\
  p(T,v_1) &=& p(T,v_2)\,.
\end{eqnarray}
Since Eq.~(\ref{eq:EOSVanDerWaals}) diverges at $v=b^{\prime}$ the
numeric solution of this equation is very challenging. In this work the
CONLES algorithm\cite{Shacham1986} was used to find the solution of this
constrained system of equations. Initial values were calculated based on
the approximative equations given in Ref.~\onlinecite{BerberanSantos2008}. For the
calculation of the specific Gibbs free energy and the specific entropy the
equations given in Ref.~\onlinecite{Younglove1994} were used which provide an excellent
approach to the calculation of these quantities based on a given equation of state
and a chosen value for the specific heat capacity at constant pressure $c_p^0(T)$
of the ideal hydrogen gas. Since the specific heat capacity has to be calculated
rather often, the data, specified in Ref.~\onlinecite{Wooley1948} as a summation,
was fitted by the following formula, which was inspired by the work presented in
Ref.~\onlinecite{Younglove1994}~and~\onlinecite{Lemmon2005}:
\begin{equation}
  \frac{c_p^0}{R_s} =
    2.5 + \sum_{l=1}^{N_{\text{e}}}
    {u_l\left(\frac{T_l}{T}\right)^2
    \frac{\exp(T_l/T)}{(1-\exp(T_l/T))^2}}\, ,
  \label{eq:FitHeatCapacity}
\end{equation}
with the parameters $u_l$ and $T_l$ given in Table~\ref{tab:FitHeatCapacity}.

\begin{table}
  \begin{ruledtabular}    
      \begin{tabular}{lrrrr}
           & \multicolumn{2}{c}{Para hydrogen} & \multicolumn{2}{c}{Normal hydrogen} \\
       $l$ & $T_l/\text{K}$ & \multicolumn{1}{c}{$u_l$} & $T_l/\text{K}$ & \multicolumn{1}{c}{$u_l$} \\ \hline
        1  &   497 &   4.18004(63) &   533 &  1.67471(33) \\
        2  &   822 &  13.997(12)   &   714 & -0.45027(49) \\
        3  &   968 & -49.866(27)   &  1908 & -0.8941(13)  \\
        4  &  1161 &  51.721(27)   &  2287 &  0.7924(13)  \\
        5  &  1336 & -19.069(11)   &  6846 &  1.4730(25)  \\
        6  &  5059 &   0.7763(13)  &       &              \\
        7  & 10247 &   2.288(39)   &       &              \\
      \end{tabular}
  \end{ruledtabular}
    \caption{Parameters of the fit function Eq.~(\ref{eq:FitHeatCapacity}) used
		  to calculate the specific heat capacity at constant pressure of
			ideal hydrogen gas. The parameters $T_l$ were chosen empirically and only
			the parameters $u_l$ were fitted to the data given in Ref.~\onlinecite{Wooley1948}.}
    \label{tab:FitHeatCapacity}
\end{table}
\begin{figure}[b]
\includegraphics[width=\columnwidth]{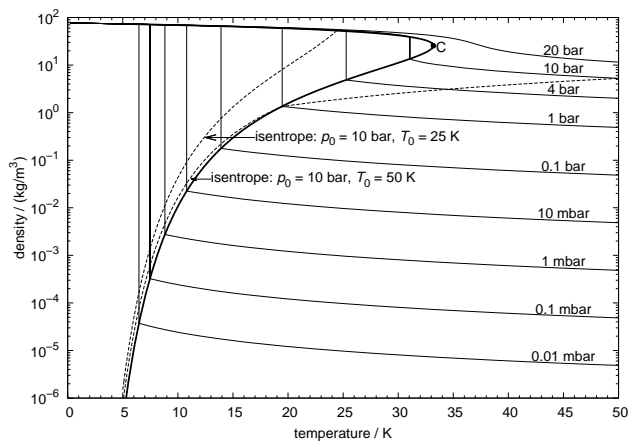}
\caption{Density of normal hydrogen as function of the temperature for
  different isobars and two selected isentropes. The boundary of the 
	coexistance region is indicated by the thick solid line.}
\label{fig:DensityVanDerWaals}
\end{figure}

In Fig.~\ref{fig:DensityVanDerWaals} the density of normal hydrogen is presented
as function of the temperature for different isobars and two selected isentropes.
The boundary of the coexistence region, where the isobars are simple vertical lines,
is indicated by the thick solid line. It is obvious that both isentropes cross this
boundary so that the calculation of the fluid properties has to be done also in
the coexistence region. For these calculations a new parameter called quality $x$
is useful, which is the ratio of the mass of the vapor phase $m_\text{g}$
and the total mass $m=m_\text{g}+m_\text{l}$ of the vapor and the liquid phase
(see, e.g., Ref.~\onlinecite{Baehr2005}):
\begin{equation}
  x = \frac{m_\text{g}}{m_\text{g}+m_\text{l}}\,.
\label{eq:Quality}
\end{equation}
This definition leads to the following equations, which connect the value of a
certain quantity $q$ with the two values of the quantity at the liquid $q_\text{l}$
and the vapor $q_\text{g}$ side of the boundary of the coexistance region
(see, e.g., Ref.~\onlinecite{Baehr2005}):
\begin{eqnarray}
  v &=& x\,v_{\text{g}} + (1-x)\,v_{\text{l}}\, , \\
  \frac{1}{\rho} &=& x\,\frac{1}{\rho_{\text{g}}} + (1-x)\,\frac{1}{\rho_{\text{l}}} \, ,\\
  h &=& x\,h_{\text{g}} + (1-x)\,h_{\text{l}}\, ,\\
  s &=& x\,s_{\text{g}} + (1-x)\,s_{\text{l}}\, ,
\end{eqnarray}
where $v$ is the specific volume, $\rho$ the density, $h$ the specific enthalpy,
and $s$ the specific entropy.
In order to calculate the specific entropy $s(T,\rho)$ the following
scheme is used:
\begin{itemize}
	\item For $T \ge T_{\text{c}}$: Calculate $s(T,\rho)$ directly from the equation
	      of state.
	\item For $T<T_{\text{c}}$: Calculate the density of the 
	  liquid phase~$\rho_{\text{l}}(T)$ and of the vapor phase~$\rho_{\text{g}}(T)$.	  
		\begin{itemize}
			\item For $\rho\le\rho_{\text{g}}$ or $\rho\ge\rho_{\text{l}}$:
			  Calculate $s(T,\rho)$ directly from the equation of state.
			\item For $\rho_{\text{g}}<\rho<\rho_{\text{l}}$:
			\begin{enumerate}
				\item Calculate the specific entropy of the liquid phase
				  $s_{\text{l}}=s(T,\rho_{\text{l}})$ and the vapor phase
					$s_{\text{g}}=s(T,\rho_{\text{g}})$.
				\item Calculate the quality $x = x(\rho,\rho_{\text{g}},\rho_{\text{l}})$.
				\item Calculate the specific entropy
				  $s = x\,s_{\text{g}} + (1-x)\,s_{\text{l}}$.
			\end{enumerate}
		\end{itemize}
\end{itemize}

In order to calculate the position dependent quantities inside of the nozzle the
velocity of sound $a$ is needed to find the critical velocity $u^*$. In the
coexistance region the following equation from Ref.~\onlinecite{VDIHeatAtlas}
is used:
  \begin{equation}
    \frac{1}{\rho\,a^2} =
      \frac{\alpha}{\rho_{\text{g}}\,a_{\text{g}}^2} +
      \frac{1-\alpha}{\rho_{\text{l}}\,a_{\text{l}}^2}\, ,
  \end{equation}
where $a_{\text{l,g}}=a(T,\rho_{\text{l,g}})$ is the velocity of sound of
the liquid and the vapor phase calculated from the equation of state
and $\alpha$ is the void fraction, defined as the ratio of the 
volume $V_{\text{g}}$ of the vapor phase to the total volume
$V_{\text{l}}+V_{\text{g}}$:
\begin{equation}
  \alpha = \frac{V_{\text{g}}}{V_{\text{l}}+V_{\text{g}}}\, .
\end{equation}


\begin{thebibliography}{32}%
\makeatletter
\providecommand \@ifxundefined [1]{%
 \@ifx{#1\undefined}
}%
\providecommand \@ifnum [1]{%
 \ifnum #1\expandafter \@firstoftwo
 \else \expandafter \@secondoftwo
 \fi
}%
\providecommand \@ifx [1]{%
 \ifx #1\expandafter \@firstoftwo
 \else \expandafter \@secondoftwo
 \fi
}%
\providecommand \natexlab [1]{#1}%
\providecommand \enquote  [1]{``#1''}%
\providecommand \bibnamefont  [1]{#1}%
\providecommand \bibfnamefont [1]{#1}%
\providecommand \citenamefont [1]{#1}%
\providecommand \href@noop [0]{\@secondoftwo}%
\providecommand \href [0]{\begingroup \@sanitize@url \@href}%
\providecommand \@href[1]{\@@startlink{#1}\@@href}%
\providecommand \@@href[1]{\endgroup#1\@@endlink}%
\providecommand \@sanitize@url [0]{\catcode `\\12\catcode `\$12\catcode
  `\&12\catcode `\#12\catcode `\^12\catcode `\_12\catcode `\%12\relax}%
\providecommand \@@startlink[1]{}%
\providecommand \@@endlink[0]{}%
\providecommand \url  [0]{\begingroup\@sanitize@url \@url }%
\providecommand \@url [1]{\endgroup\@href {#1}{\urlprefix }}%
\providecommand \urlprefix  [0]{URL }%
\providecommand \Eprint [0]{\href }%
\providecommand \doibase [0]{http://dx.doi.org/}%
\providecommand \selectlanguage [0]{\@gobble}%
\providecommand \bibinfo  [0]{\@secondoftwo}%
\providecommand \bibfield  [0]{\@secondoftwo}%
\providecommand \translation [1]{[#1]}%
\providecommand \BibitemOpen [0]{}%
\providecommand \bibitemStop [0]{}%
\providecommand \bibitemNoStop [0]{.\EOS\space}%
\providecommand \EOS [0]{\spacefactor3000\relax}%
\providecommand \BibitemShut  [1]{\csname bibitem#1\endcsname}%
\let\auto@bib@innerbib\@empty
\bibitem [{\citenamefont {Pauly}(2000)}]{Pauly2000}%
  \BibitemOpen
  \bibfield  {author} {\bibinfo {author} {\bibfnamefont {H.}~\bibnamefont
  {Pauly}},\ }\href@noop {} {\emph {\bibinfo {title} {{Atom, Molecule, and
  Cluster Beams II - Cluster Beams, Fast and Slow Beams, Accessory Equipment
  and Applications}}}}\ (\bibinfo  {publisher} {Springer Berlin Heidelberg},\
  \bibinfo {year} {2000})\BibitemShut {NoStop}%
\bibitem [{\citenamefont {Barsov}\ \emph {et~al.}(2001)\citenamefont {Barsov},
  \citenamefont {Bechstedt}, \citenamefont {Bothe}, \citenamefont {Bongers},
  \citenamefont {Borchert}, \citenamefont {Borgs}, \citenamefont
  {Br\"{a}utigam}, \citenamefont {B\"{u}scher}, \citenamefont {Cassing},
  \citenamefont {Chernyshev}, \citenamefont {Chiladze}, \citenamefont
  {Dietrich}, \citenamefont {Drochner}, \citenamefont {Dymov}, \citenamefont
  {Erven}, \citenamefont {Esser}, \citenamefont {Franzen}, \citenamefont
  {Golubeva}, \citenamefont {Gotta}, \citenamefont {Grande}, \citenamefont
  {Grzonka}, \citenamefont {Hardt}, \citenamefont {Hartmann}, \citenamefont
  {Hejny}, \citenamefont {von Horn}, \citenamefont {Jarczyk}, \citenamefont
  {Junghans}, \citenamefont {Kacharava}, \citenamefont {Kamys}, \citenamefont
  {Khoukaz}, \citenamefont {Kirchner}, \citenamefont {Klehr}, \citenamefont
  {Klein}, \citenamefont {Koch}, \citenamefont {Komarov}, \citenamefont
  {Kondratyuk}, \citenamefont {Koptev}, \citenamefont {Kopyto}, \citenamefont
  {Krause}, \citenamefont {Kravtsov}, \citenamefont {Kruglov}, \citenamefont
  {Kulessa}, \citenamefont {Kulikov}, \citenamefont {Lang}, \citenamefont
  {Langenhagen}, \citenamefont {Lepges}, \citenamefont {Ley}, \citenamefont
  {Maier}, \citenamefont {Martin}, \citenamefont {Macharashvili}, \citenamefont
  {Merzliakov}, \citenamefont {Meyer}, \citenamefont {Mikirtychiants},
  \citenamefont {M\"{u}ller}, \citenamefont {Munhofen}, \citenamefont
  {Mussgiller}, \citenamefont {Nekipelov}, \citenamefont {Nelyubin},
  \citenamefont {Nioradze}, \citenamefont {Ohm}, \citenamefont {Petrus},
  \citenamefont {Prasuhn}, \citenamefont {Prietzschk}, \citenamefont {Probst},
  \citenamefont {Pysz}, \citenamefont {Rathmann}, \citenamefont {Rimarzig},
  \citenamefont {Rudy}, \citenamefont {Santo}, \citenamefont {gen. Schieck},
  \citenamefont {Schleichert}, \citenamefont {Schneider}, \citenamefont
  {Schneider}, \citenamefont {Schneider}, \citenamefont {Schwarz},
  \citenamefont {Seyfarth}, \citenamefont {Sibirtsev}, \citenamefont {Sieling},
  \citenamefont {Sistemich}, \citenamefont {Selikov}, \citenamefont
  {Stechemesser}, \citenamefont {Stein}, \citenamefont {Strzalkowski},
  \citenamefont {Watzlawik}, \citenamefont {W\"{u}stner}, \citenamefont
  {Yashenko}, \citenamefont {Zalikhanov}, \citenamefont {Zhuravlev},
  \citenamefont {Zwoll}, \citenamefont {Zychor}, \citenamefont {Schult},\ and\
  \citenamefont {Str\"{o}her}}]{Barsov2001}%
  \BibitemOpen
  \bibfield  {author} {\bibinfo {author} {\bibfnamefont {S.}~\bibnamefont
  {Barsov}}, \bibinfo {author} {\bibfnamefont {U.}~\bibnamefont {Bechstedt}},
  \bibinfo {author} {\bibfnamefont {W.}~\bibnamefont {Bothe}}, \bibinfo
  {author} {\bibfnamefont {N.}~\bibnamefont {Bongers}}, \bibinfo {author}
  {\bibfnamefont {G.}~\bibnamefont {Borchert}}, \bibinfo {author}
  {\bibfnamefont {W.}~\bibnamefont {Borgs}}, \bibinfo {author} {\bibfnamefont
  {W.}~\bibnamefont {Br\"{a}utigam}}, \bibinfo {author} {\bibfnamefont
  {M.}~\bibnamefont {B\"{u}scher}}, \bibinfo {author} {\bibfnamefont
  {W.}~\bibnamefont {Cassing}}, \bibinfo {author} {\bibfnamefont
  {V.}~\bibnamefont {Chernyshev}}, \bibinfo {author} {\bibfnamefont
  {B.}~\bibnamefont {Chiladze}}, \bibinfo {author} {\bibfnamefont
  {J.}~\bibnamefont {Dietrich}}, \bibinfo {author} {\bibfnamefont
  {M.}~\bibnamefont {Drochner}}, \bibinfo {author} {\bibfnamefont
  {S.}~\bibnamefont {Dymov}}, \bibinfo {author} {\bibfnamefont
  {W.}~\bibnamefont {Erven}}, \bibinfo {author} {\bibfnamefont
  {R.}~\bibnamefont {Esser}}, \bibinfo {author} {\bibfnamefont
  {A.}~\bibnamefont {Franzen}}, \bibinfo {author} {\bibfnamefont
  {Y.}~\bibnamefont {Golubeva}}, \bibinfo {author} {\bibfnamefont
  {D.}~\bibnamefont {Gotta}}, \bibinfo {author} {\bibfnamefont
  {T.}~\bibnamefont {Grande}}, \bibinfo {author} {\bibfnamefont
  {D.}~\bibnamefont {Grzonka}}, \bibinfo {author} {\bibfnamefont
  {A.}~\bibnamefont {Hardt}}, \bibinfo {author} {\bibfnamefont
  {M.}~\bibnamefont {Hartmann}}, \bibinfo {author} {\bibfnamefont
  {V.}~\bibnamefont {Hejny}}, \bibinfo {author} {\bibfnamefont
  {L.}~\bibnamefont {von Horn}}, \bibinfo {author} {\bibfnamefont
  {L.}~\bibnamefont {Jarczyk}}, \bibinfo {author} {\bibfnamefont
  {H.}~\bibnamefont {Junghans}}, \bibinfo {author} {\bibfnamefont
  {A.}~\bibnamefont {Kacharava}}, \bibinfo {author} {\bibfnamefont
  {B.}~\bibnamefont {Kamys}}, \bibinfo {author} {\bibfnamefont
  {A.}~\bibnamefont {Khoukaz}}, \bibinfo {author} {\bibfnamefont
  {T.}~\bibnamefont {Kirchner}}, \bibinfo {author} {\bibfnamefont
  {F.}~\bibnamefont {Klehr}}, \bibinfo {author} {\bibfnamefont
  {W.}~\bibnamefont {Klein}}, \bibinfo {author} {\bibfnamefont {H.~R.}\
  \bibnamefont {Koch}}, \bibinfo {author} {\bibfnamefont {V.~I.}\ \bibnamefont
  {Komarov}}, \bibinfo {author} {\bibfnamefont {L.}~\bibnamefont {Kondratyuk}},
  \bibinfo {author} {\bibfnamefont {V.}~\bibnamefont {Koptev}}, \bibinfo
  {author} {\bibfnamefont {S.}~\bibnamefont {Kopyto}}, \bibinfo {author}
  {\bibfnamefont {R.}~\bibnamefont {Krause}}, \bibinfo {author} {\bibfnamefont
  {P.}~\bibnamefont {Kravtsov}}, \bibinfo {author} {\bibfnamefont
  {V.}~\bibnamefont {Kruglov}}, \bibinfo {author} {\bibfnamefont
  {P.}~\bibnamefont {Kulessa}}, \bibinfo {author} {\bibfnamefont
  {A.}~\bibnamefont {Kulikov}}, \bibinfo {author} {\bibfnamefont
  {N.}~\bibnamefont {Lang}}, \bibinfo {author} {\bibfnamefont {N.}~\bibnamefont
  {Langenhagen}}, \bibinfo {author} {\bibfnamefont {A.}~\bibnamefont {Lepges}},
  \bibinfo {author} {\bibfnamefont {J.}~\bibnamefont {Ley}}, \bibinfo {author}
  {\bibfnamefont {R.}~\bibnamefont {Maier}}, \bibinfo {author} {\bibfnamefont
  {S.}~\bibnamefont {Martin}}, \bibinfo {author} {\bibfnamefont
  {G.}~\bibnamefont {Macharashvili}}, \bibinfo {author} {\bibfnamefont
  {S.}~\bibnamefont {Merzliakov}}, \bibinfo {author} {\bibfnamefont
  {K.}~\bibnamefont {Meyer}}, \bibinfo {author} {\bibfnamefont
  {S.}~\bibnamefont {Mikirtychiants}}, \bibinfo {author} {\bibfnamefont
  {H.}~\bibnamefont {M\"{u}ller}}, \bibinfo {author} {\bibfnamefont
  {P.}~\bibnamefont {Munhofen}}, \bibinfo {author} {\bibfnamefont
  {A.}~\bibnamefont {Mussgiller}}, \bibinfo {author} {\bibfnamefont
  {M.}~\bibnamefont {Nekipelov}}, \bibinfo {author} {\bibfnamefont
  {V.}~\bibnamefont {Nelyubin}}, \bibinfo {author} {\bibfnamefont
  {M.}~\bibnamefont {Nioradze}}, \bibinfo {author} {\bibfnamefont
  {H.}~\bibnamefont {Ohm}}, \bibinfo {author} {\bibfnamefont {A.}~\bibnamefont
  {Petrus}}, \bibinfo {author} {\bibfnamefont {D.}~\bibnamefont {Prasuhn}},
  \bibinfo {author} {\bibfnamefont {B.}~\bibnamefont {Prietzschk}}, \bibinfo
  {author} {\bibfnamefont {H.~J.}\ \bibnamefont {Probst}}, \bibinfo {author}
  {\bibfnamefont {K.}~\bibnamefont {Pysz}}, \bibinfo {author} {\bibfnamefont
  {F.}~\bibnamefont {Rathmann}}, \bibinfo {author} {\bibfnamefont
  {B.}~\bibnamefont {Rimarzig}}, \bibinfo {author} {\bibfnamefont
  {Z.}~\bibnamefont {Rudy}}, \bibinfo {author} {\bibfnamefont {R.}~\bibnamefont
  {Santo}}, \bibinfo {author} {\bibfnamefont {H.}\ \bibnamefont {Paetz gen.
  Schieck}}, \bibinfo {author} {\bibfnamefont {R.}~\bibnamefont {Schleichert}},
  \bibinfo {author} {\bibfnamefont {A.}~\bibnamefont {Schneider}}, \bibinfo
  {author} {\bibfnamefont {C.}~\bibnamefont {Schneider}}, \bibinfo {author}
  {\bibfnamefont {H.}~\bibnamefont {Schneider}}, \bibinfo {author}
  {\bibfnamefont {U.}~\bibnamefont {Schwarz}}, \bibinfo {author} {\bibfnamefont
  {H.}~\bibnamefont {Seyfarth}}, \bibinfo {author} {\bibfnamefont
  {A.}~\bibnamefont {Sibirtsev}}, \bibinfo {author} {\bibfnamefont
  {U.}~\bibnamefont {Sieling}}, \bibinfo {author} {\bibfnamefont
  {K.}~\bibnamefont {Sistemich}}, \bibinfo {author} {\bibfnamefont
  {A.}~\bibnamefont {Selikov}}, \bibinfo {author} {\bibfnamefont
  {H.}~\bibnamefont {Stechemesser}}, \bibinfo {author} {\bibfnamefont {H.~J.}\
  \bibnamefont {Stein}}, \bibinfo {author} {\bibfnamefont {A.}~\bibnamefont
  {Strzalkowski}}, \bibinfo {author} {\bibfnamefont {K.-H.}\ \bibnamefont
  {Watzlawik}}, \bibinfo {author} {\bibfnamefont {P.}~\bibnamefont
  {W\"{u}stner}}, \bibinfo {author} {\bibfnamefont {S.}~\bibnamefont
  {Yashenko}}, \bibinfo {author} {\bibfnamefont {B.}~\bibnamefont
  {Zalikhanov}}, \bibinfo {author} {\bibfnamefont {N.}~\bibnamefont
  {Zhuravlev}}, \bibinfo {author} {\bibfnamefont {K.}~\bibnamefont {Zwoll}},
  \bibinfo {author} {\bibfnamefont {I.}~\bibnamefont {Zychor}}, \bibinfo
  {author} {\bibfnamefont {O.~W.~B.}\ \bibnamefont {Schult}}, \ and\ \bibinfo
  {author} {\bibfnamefont {H.}~\bibnamefont {Str\"{o}her}},\ }\href@noop {}
  {\bibfield  {journal} {\bibinfo  {journal} {Nucl. Instrum. Methods A}\
  }\textbf {\bibinfo {volume} {462}},\ \bibinfo {pages} {364 } (\bibinfo {year}
  {2001})}\BibitemShut {NoStop}%
\bibitem [{\citenamefont {Brauksiepe}\ \emph {et~al.}(1996)\citenamefont
  {Brauksiepe}, \citenamefont {Grzonka}, \citenamefont {Kilian}, \citenamefont
  {Oelert}, \citenamefont {Roderburg}, \citenamefont {Rook}, \citenamefont
  {Sefzick}, \citenamefont {Turek}, \citenamefont {Wolke}, \citenamefont
  {Bechstedt}, \citenamefont {Dietrich}, \citenamefont {Maier}, \citenamefont
  {Martin}, \citenamefont {Prasuhn}, \citenamefont {Schnase}, \citenamefont
  {Schneider}, \citenamefont {Stockhorst}, \citenamefont {T\"{o}lle},
  \citenamefont {Karnadi}, \citenamefont {Nellen}, \citenamefont {Watzlawik},
  \citenamefont {Diart}, \citenamefont {Gutschmidt}, \citenamefont {Jochmann},
  \citenamefont {K\"{o}hler}, \citenamefont {Reinartz}, \citenamefont
  {W\"{u}stner}, \citenamefont {Zwoll}, \citenamefont {Klehr}, \citenamefont
  {Stechemesser}, \citenamefont {Dombrowski}, \citenamefont {Hamsink},
  \citenamefont {Khoukaz}, \citenamefont {Lister}, \citenamefont {Quentmeier},
  \citenamefont {Santo}, \citenamefont {Schepers}, \citenamefont {Jarczyk},
  \citenamefont {Kozela}, \citenamefont {Majewski}, \citenamefont {Misiak},
  \citenamefont {Moskal}, \citenamefont {Smyrski}, \citenamefont {Sokolowski},
  \citenamefont {Strzalkowski}, \citenamefont {Balewski}, \citenamefont
  {Budzanowski}, \citenamefont {Bowes}, \citenamefont {Hardt}, \citenamefont
  {Goodman}, \citenamefont {Seddik},\ and\ \citenamefont
  {Ziolkowski}}]{Brauksiepe1996}%
  \BibitemOpen
  \bibfield  {author} {\bibinfo {author} {\bibfnamefont {S.}~\bibnamefont
  {Brauksiepe}}, \bibinfo {author} {\bibfnamefont {D.}~\bibnamefont {Grzonka}},
  \bibinfo {author} {\bibfnamefont {K.}~\bibnamefont {Kilian}}, \bibinfo
  {author} {\bibfnamefont {W.}~\bibnamefont {Oelert}}, \bibinfo {author}
  {\bibfnamefont {E.}~\bibnamefont {Roderburg}}, \bibinfo {author}
  {\bibfnamefont {M.}~\bibnamefont {Rook}}, \bibinfo {author} {\bibfnamefont
  {T.}~\bibnamefont {Sefzick}}, \bibinfo {author} {\bibfnamefont
  {P.}~\bibnamefont {Turek}}, \bibinfo {author} {\bibfnamefont
  {M.}~\bibnamefont {Wolke}}, \bibinfo {author} {\bibfnamefont
  {U.}~\bibnamefont {Bechstedt}}, \bibinfo {author} {\bibfnamefont
  {J.}~\bibnamefont {Dietrich}}, \bibinfo {author} {\bibfnamefont
  {R.}~\bibnamefont {Maier}}, \bibinfo {author} {\bibfnamefont
  {S.}~\bibnamefont {Martin}}, \bibinfo {author} {\bibfnamefont
  {D.}~\bibnamefont {Prasuhn}}, \bibinfo {author} {\bibfnamefont
  {A.}~\bibnamefont {Schnase}}, \bibinfo {author} {\bibfnamefont
  {H.}~\bibnamefont {Schneider}}, \bibinfo {author} {\bibfnamefont
  {H.}~\bibnamefont {Stockhorst}}, \bibinfo {author} {\bibfnamefont
  {R.}~\bibnamefont {T\"{o}lle}}, \bibinfo {author} {\bibfnamefont
  {M.}~\bibnamefont {Karnadi}}, \bibinfo {author} {\bibfnamefont
  {R.}~\bibnamefont {Nellen}}, \bibinfo {author} {\bibfnamefont
  {K.}~\bibnamefont {Watzlawik}}, \bibinfo {author} {\bibfnamefont
  {K.}~\bibnamefont {Diart}}, \bibinfo {author} {\bibfnamefont
  {H.}~\bibnamefont {Gutschmidt}}, \bibinfo {author} {\bibfnamefont
  {M.}~\bibnamefont {Jochmann}}, \bibinfo {author} {\bibfnamefont
  {M.}~\bibnamefont {K\"{o}hler}}, \bibinfo {author} {\bibfnamefont
  {R.}~\bibnamefont {Reinartz}}, \bibinfo {author} {\bibfnamefont
  {P.}~\bibnamefont {W\"{u}stner}}, \bibinfo {author} {\bibfnamefont
  {K.}~\bibnamefont {Zwoll}}, \bibinfo {author} {\bibfnamefont
  {F.}~\bibnamefont {Klehr}}, \bibinfo {author} {\bibfnamefont
  {H.}~\bibnamefont {Stechemesser}}, \bibinfo {author} {\bibfnamefont
  {H.}~\bibnamefont {Dombrowski}}, \bibinfo {author} {\bibfnamefont
  {W.}~\bibnamefont {Hamsink}}, \bibinfo {author} {\bibfnamefont
  {A.}~\bibnamefont {Khoukaz}}, \bibinfo {author} {\bibfnamefont
  {T.}~\bibnamefont {Lister}}, \bibinfo {author} {\bibfnamefont
  {C.}~\bibnamefont {Quentmeier}}, \bibinfo {author} {\bibfnamefont
  {R.}~\bibnamefont {Santo}}, \bibinfo {author} {\bibfnamefont
  {G.}~\bibnamefont {Schepers}}, \bibinfo {author} {\bibfnamefont
  {L.}~\bibnamefont {Jarczyk}}, \bibinfo {author} {\bibfnamefont
  {A.}~\bibnamefont {Kozela}}, \bibinfo {author} {\bibfnamefont
  {J.}~\bibnamefont {Majewski}}, \bibinfo {author} {\bibfnamefont
  {A.}~\bibnamefont {Misiak}}, \bibinfo {author} {\bibfnamefont
  {P.}~\bibnamefont {Moskal}}, \bibinfo {author} {\bibfnamefont
  {J.}~\bibnamefont {Smyrski}}, \bibinfo {author} {\bibfnamefont
  {M.}~\bibnamefont {Sokolowski}}, \bibinfo {author} {\bibfnamefont
  {A.}~\bibnamefont {Strzalkowski}}, \bibinfo {author} {\bibfnamefont
  {J.}~\bibnamefont {Balewski}}, \bibinfo {author} {\bibfnamefont
  {A.}~\bibnamefont {Budzanowski}}, \bibinfo {author} {\bibfnamefont
  {S.}~\bibnamefont {Bowes}}, \bibinfo {author} {\bibfnamefont
  {A.}~\bibnamefont {Hardt}}, \bibinfo {author} {\bibfnamefont
  {C.}~\bibnamefont {Goodman}}, \bibinfo {author} {\bibfnamefont
  {U.}~\bibnamefont {Seddik}}, \ and\ \bibinfo {author} {\bibfnamefont
  {M.}~\bibnamefont {Ziolkowski}},\ }\href@noop {} {\bibfield  {journal}
  {\bibinfo  {journal} {Nucl. Instrum. Methods A}\ }\textbf {\bibinfo {volume}
  {376}},\ \bibinfo {pages} {397} (\bibinfo {year} {1996})}\BibitemShut
  {NoStop}%
\bibitem [{\citenamefont {Maier}(1997)}]{Maier1997}%
  \BibitemOpen
  \bibfield  {author} {\bibinfo {author} {\bibfnamefont {R.}~\bibnamefont
  {Maier}},\ }\href@noop {} {\bibfield  {journal} {\bibinfo  {journal} {Nucl.
  Instrum. Methods A}\ }\textbf {\bibinfo {volume} {390}},\ \bibinfo {pages}
  {1} (\bibinfo {year} {1997})}\BibitemShut {NoStop}%
\bibitem [{\citenamefont {{PANDA Collaboration}}(2005)}]{PANDA2005}%
  \BibitemOpen
  \bibfield  {author} {\bibinfo {author} {\bibnamefont {{PANDA
  Collaboration}}},\ }\href@noop {} {\enquote {\bibinfo {title} {Strong
  interaction studies with antiprotons}}},\ {\bibinfo {type}
  {{Technical Progress Report}}} (\bibinfo  {publisher} {FAIR GmbH},\ \bibinfo {year}
  {2005})\BibitemShut {NoStop}%
\bibitem [{\citenamefont {T\"{a}schner}\ \emph {et~al.}(2011)\citenamefont
  {T\"{a}schner}, \citenamefont {K\"{o}hler}, \citenamefont {Ortjohann},\ and\
  \citenamefont {Khoukaz}}]{Taeschner2011}%
  \BibitemOpen
  \bibfield  {author} {\bibinfo {author} {\bibfnamefont {A.}~\bibnamefont
  {T\"{a}schner}}, \bibinfo {author} {\bibfnamefont {E.}~\bibnamefont
  {K\"{o}hler}}, \bibinfo {author} {\bibfnamefont {H.-W.}\ \bibnamefont
  {Ortjohann}}, \ and\ \bibinfo {author} {\bibfnamefont {A.}~\bibnamefont
  {Khoukaz}},\ }\href {\doibase 10.1016/j.nima.2011.09.024} {\bibfield
  {journal} {\bibinfo  {journal} {Nucl. Instrum. Methods A}\ }\textbf {\bibinfo
  {volume} {660}},\ \bibinfo {pages} {22} (\bibinfo {year} {2011})},\ \Eprint
  {http://arxiv.org/abs/1108.2653} {arXiv:1108.2653 [physics.ins-det]}
  \BibitemShut {NoStop}%
\bibitem [{\citenamefont {T\"{a}schner}(2013)}]{Taeschner2013}%
  \BibitemOpen
  \bibfield  {author} {\bibinfo {author} {\bibfnamefont {A.}~\bibnamefont
  {T\"{a}schner}},\ }\emph {\bibinfo {title} {Entwicklung und Untersuchung von
  Cluster-Jet-Targets h\"{o}chster Dichte}},\ \href@noop {} {\bibinfo {type}
  {{Doctoral Thesis}}},\ \bibinfo  {school} {Westf\"{a}lische
  Wilhems-Universit\"{a}t, M\"{u}nster} (\bibinfo {year} {2013})\BibitemShut
  {NoStop}%
\bibitem [{\citenamefont {Hinterberger}(2008)}]{Hinterberger2008}%
  \BibitemOpen
  \bibfield  {author} {\bibinfo {author} {\bibfnamefont {F.}~\bibnamefont
  {Hinterberger}},\ }\href@noop {} {\emph {\bibinfo {title} {Physik der
  Teilchenbeschleuniger und Ionenoptik}}}\ (\bibinfo  {publisher} {Springer
  Berlin Heidelberg},\ \bibinfo {year} {2008})\BibitemShut {NoStop}%
\bibitem [{\citenamefont {Knuth}, \citenamefont {Schunemann},\ and\
  \citenamefont {Toennies}(1995)}]{Knuth1995}%
  \BibitemOpen
  \bibfield  {author} {\bibinfo {author} {\bibfnamefont {E.}~\bibnamefont
  {Knuth}}, \bibinfo {author} {\bibfnamefont {F.}~\bibnamefont {Schunemann}}, \
  and\ \bibinfo {author} {\bibfnamefont {J.~P.}\ \bibnamefont {Toennies}},\
  }\href {\doibase 10.1063/1.469072} {\bibfield  {journal} {\bibinfo  {journal}
  {J. Chem. Phys.}\ }\textbf {\bibinfo {volume} {102}},\ \bibinfo {pages}
  {6258} (\bibinfo {year} {1995})}\BibitemShut {NoStop}%
\bibitem [{\citenamefont {Allspach}\ \emph {et~al.}(1998)\citenamefont
  {Allspach}, \citenamefont {Hahn}, \citenamefont {Kendziora}, \citenamefont
  {Pordes}, \citenamefont {Boero}, \citenamefont {Garzoglio}, \citenamefont
  {Macri}, \citenamefont {Marinelli}, \citenamefont {Pallavicini},\ and\
  \citenamefont {Robutti}}]{Allspach1998}%
  \BibitemOpen
  \bibfield  {author} {\bibinfo {author} {\bibfnamefont {D.}~\bibnamefont
  {Allspach}}, \bibinfo {author} {\bibfnamefont {A.}~\bibnamefont {Hahn}},
  \bibinfo {author} {\bibfnamefont {C.}~\bibnamefont {Kendziora}}, \bibinfo
  {author} {\bibfnamefont {S.}~\bibnamefont {Pordes}}, \bibinfo {author}
  {\bibfnamefont {G.}~\bibnamefont {Boero}}, \bibinfo {author} {\bibfnamefont
  {G.}~\bibnamefont {Garzoglio}}, \bibinfo {author} {\bibfnamefont
  {M.}~\bibnamefont {Macri}}, \bibinfo {author} {\bibfnamefont
  {M.}~\bibnamefont {Marinelli}}, \bibinfo {author} {\bibfnamefont
  {M.}~\bibnamefont {Pallavicini}}, \ and\ \bibinfo {author} {\bibfnamefont
  {E.}~\bibnamefont {Robutti}},\ }\href {\doibase
  10.1016/S0168-9002(98)00236-8} {\bibfield  {journal} {\bibinfo  {journal}
  {Nucl. Instrum. Methods A}\ }\textbf {\bibinfo {volume} {410}},\ \bibinfo
  {pages} {195} (\bibinfo {year} {1998})}\BibitemShut {NoStop}%
\bibitem [{\citenamefont {Christen}, \citenamefont {Rademann},\ and\
  \citenamefont {Even}(2010)}]{Christen2010}%
  \BibitemOpen
  \bibfield  {author} {\bibinfo {author} {\bibfnamefont {W.}~\bibnamefont
  {Christen}}, \bibinfo {author} {\bibfnamefont {K.}~\bibnamefont {Rademann}},
  \ and\ \bibinfo {author} {\bibfnamefont {U.}~\bibnamefont {Even}},\ }\href
  {\doibase 10.1021/jp102855m} {\bibfield  {journal} {\bibinfo  {journal} {J.
  Phys. Chem. A}\ }\textbf {\bibinfo {volume} {114}},\ \bibinfo {pages} {11189}
  (\bibinfo {year} {2010})},\ \Eprint
  {http://arxiv.org/abs/http://pubs.acs.org/doi/pdf/10.1021/jp102855m}
  {http://pubs.acs.org/doi/pdf/10.1021/jp102855m} \BibitemShut {NoStop}%
\bibitem [{\citenamefont {T\"{a}schner}\ \emph {et~al.}(2007)\citenamefont
  {T\"{a}schner}, \citenamefont {General}, \citenamefont {Otte}, \citenamefont
  {Rausmann},\ and\ \citenamefont {Khoukaz}}]{Taeschner2007}%
  \BibitemOpen
  \bibfield  {author} {\bibinfo {author} {\bibfnamefont {A.}~\bibnamefont
  {T\"{a}schner}}, \bibinfo {author} {\bibfnamefont {S.}~\bibnamefont
  {General}}, \bibinfo {author} {\bibfnamefont {J.}~\bibnamefont {Otte}},
  \bibinfo {author} {\bibfnamefont {T.}~\bibnamefont {Rausmann}}, \ and\
  \bibinfo {author} {\bibfnamefont {A.}~\bibnamefont {Khoukaz}},\ }\href
  {\doibase 10.1063/1.2819057} {\bibfield  {journal} {\bibinfo  {journal} {AIP
  Conf.Proc.}\ }\textbf {\bibinfo {volume} {950}},\ \bibinfo {pages} {85}
  (\bibinfo {year} {2007})}\BibitemShut {NoStop}%
\bibitem [{\citenamefont {Buchenau}\ \emph {et~al.}(1990)\citenamefont
  {Buchenau}, \citenamefont {Knuth}, \citenamefont {Northby}, \citenamefont
  {Toennies},\ and\ \citenamefont {Winkler}}]{Buchenau1990}%
  \BibitemOpen
  \bibfield  {author} {\bibinfo {author} {\bibfnamefont {H.}~\bibnamefont
  {Buchenau}}, \bibinfo {author} {\bibfnamefont {E.~L.}\ \bibnamefont {Knuth}},
  \bibinfo {author} {\bibfnamefont {J.}~\bibnamefont {Northby}}, \bibinfo
  {author} {\bibfnamefont {J.~P.}\ \bibnamefont {Toennies}}, \ and\ \bibinfo
  {author} {\bibfnamefont {C.}~\bibnamefont {Winkler}},\ }\href@noop {}
  {\bibfield  {journal} {\bibinfo  {journal} {J. Chem. Phys.}\ }\textbf
  {\bibinfo {volume} {92}},\ \bibinfo {pages} {6875} (\bibinfo {year}
  {1990})}\BibitemShut {NoStop}%
\bibitem [{\citenamefont {Christen}(2013)}]{Christen2013}%
  \BibitemOpen
  \bibfield  {author} {\bibinfo {author} {\bibfnamefont {W.}~\bibnamefont
  {Christen}},\ }\href {\doibase 10.1063/1.4812371} {\bibfield  {journal}
  {\bibinfo  {journal} {J. Chem. Phys.}\ }\textbf {\bibinfo {volume} {139}},\
  \bibinfo {eid} {024202} (\bibinfo {year} {2013})}\BibitemShut {NoStop}%
\bibitem [{\citenamefont {Harms}, \citenamefont {Toennies},\ and\ \citenamefont
  {Knuth}(1997)}]{Harms1997}%
  \BibitemOpen
  \bibfield  {author} {\bibinfo {author} {\bibfnamefont {J.}~\bibnamefont
  {Harms}}, \bibinfo {author} {\bibfnamefont {J.~P.}\ \bibnamefont {Toennies}},
  \ and\ \bibinfo {author} {\bibfnamefont {E.~L.}\ \bibnamefont {Knuth}},\
  }\href {\doibase 10.1063/1.473083} {\bibfield  {journal} {\bibinfo  {journal}
  {J. Chem. Phys.}\ }\textbf {\bibinfo {volume} {106}},\ \bibinfo {pages}
  {3348} (\bibinfo {year} {1997})}\BibitemShut {NoStop}%
\bibitem [{\citenamefont {K\"{o}hler}(2010)}]{Koehler2010}%
  \BibitemOpen
  \bibfield  {author} {\bibinfo {author} {\bibfnamefont {E.~D.}\ \bibnamefont
  {K\"{o}hler}},\ }\emph {\bibinfo {title} {{Das M\"{u}nsteraner Cluster-Jet
  Target MCT2, ein Prototyp f\"{u}r das PANDA-Experiment, \& die Analyse der
  Eigenschaften des Clusterstrahls}}},\ \href@noop {} {\bibinfo {type}
  {{Diploma Thesis}}},\ \bibinfo  {school} {Westf\"{a}lische
  Wilhems-Universit\"{a}t, M\"{u}nster} (\bibinfo {year} {2010})\BibitemShut
  {NoStop}%
\bibitem [{\citenamefont {Anderson}(1990)}]{Anderson1990}%
  \BibitemOpen
  \bibfield  {author} {\bibinfo {author} {\bibfnamefont {J.~D.}\ \bibnamefont
  {Anderson}},\ }\href@noop {} {\emph {\bibinfo {title} {Modern Compressible
  Flow: With Historical Perspective}}},\ \bibinfo {edition} {international 2
  revised}\ ed.\ (\bibinfo  {publisher} {McGraw-Hill Education (ISE
  Editions)},\ \bibinfo {year} {1990})\BibitemShut {NoStop}%
\bibitem [{\citenamefont {McCarty}, \citenamefont {Hord},\ and\ \citenamefont
  {Rode}(1981)}]{McCarty1981}%
  \BibitemOpen
  \bibfield  {author} {\bibinfo {author} {\bibfnamefont {R.~D.}\ \bibnamefont
  {McCarty}}, \bibinfo {author} {\bibfnamefont {J.}~\bibnamefont {Hord}}, \
  and\ \bibinfo {author} {\bibfnamefont {H.~M.}\ \bibnamefont {Rode}},\
  }\href@noop {} {\enquote {\bibinfo {title} {Selected properties of hydrogen
  (engineering design data)},}\ }\bibinfo {type} {Monograph}\ \bibinfo {number}
  {168}\ (\bibinfo  {institution} {National Bureau of Standards, Washington,
  DC},\ \bibinfo {year} {1981})\BibitemShut {NoStop}%
\bibitem [{\citenamefont {Leachman}\ \emph {et~al.}(2009)\citenamefont
  {Leachman}, \citenamefont {Jacobsen}, \citenamefont {Penoncello},\ and\
  \citenamefont {Lemmon}}]{Leachman2009}%
  \BibitemOpen
  \bibfield  {author} {\bibinfo {author} {\bibfnamefont {J.~W.}\ \bibnamefont
  {Leachman}}, \bibinfo {author} {\bibfnamefont {R.~T.}\ \bibnamefont
  {Jacobsen}}, \bibinfo {author} {\bibfnamefont {S.~G.}\ \bibnamefont
  {Penoncello}}, \ and\ \bibinfo {author} {\bibfnamefont {E.~W.}\ \bibnamefont
  {Lemmon}},\ }\href {\doibase 10.1063/1.3160306} {\bibfield  {journal}
  {\bibinfo  {journal} {J. Phys. Chem. Ref. Data}\ }\textbf {\bibinfo {volume}
  {38}},\ \bibinfo {pages} {721} (\bibinfo {year} {2009})}\BibitemShut
  {NoStop}%
\bibitem [{\citenamefont {Ekstr\"{o}m}(1995)}]{Ekstroem1995}%
  \BibitemOpen
  \bibfield  {author} {\bibinfo {author} {\bibfnamefont {C.}~\bibnamefont
  {Ekstr\"{o}m}},\ }\href {\doibase DOI: 10.1016/0168-9002(95)00240-5}
  {\bibfield  {journal} {\bibinfo  {journal} {Nucl. Instrum. Methods A}\
  }\textbf {\bibinfo {volume} {362}},\ \bibinfo {pages} {1} (\bibinfo {year}
  {1995})},\ \bibinfo {note} {proceedings of the 17th World Conference of the
  International Nuclear Target Development Society}\BibitemShut {NoStop}%
\bibitem [{\citenamefont {Nolting}(1997)}]{Nolting4}%
  \BibitemOpen
  \bibfield  {author} {\bibinfo {author} {\bibfnamefont {W.}~\bibnamefont
  {Nolting}},\ }\href@noop {} {\emph {\bibinfo {title} {Grundkurs Theoretische
  Physik, 4, Spezielle Relativit\"{a}tstheorie, Thermodynamic}}}\ (\bibinfo
  {publisher} {Vieweg},\ \bibinfo {year} {1997})\BibitemShut {NoStop}%
\bibitem [{\citenamefont {Sedunov}(2011)}]{Sedunov2011}%
  \BibitemOpen
  \bibfield  {author} {\bibinfo {author} {\bibfnamefont {B.~I.}\ \bibnamefont
  {Sedunov}},\ }\href {\doibase 10.1155/2011/194353} {\bibfield  {journal}
  {\bibinfo  {journal} {Journal of Thermodynamics}\ }\textbf {\bibinfo {volume}
  {2011}},\ \bibinfo {pages} {1} (\bibinfo {year} {2011})}\BibitemShut
  {NoStop}%
\bibitem [{\citenamefont {Sallet}(1991)}]{Sallet1991}%
  \BibitemOpen
  \bibfield  {author} {\bibinfo {author} {\bibfnamefont {D.~W.}\ \bibnamefont
  {Sallet}},\ }\href@noop {} {\bibfield  {journal} {\bibinfo  {journal} {Heat
  and Mass Transfer}\ }\textbf {\bibinfo {volume} {26}},\ \bibinfo {pages}
  {315} (\bibinfo {year} {1991})}\BibitemShut {NoStop}%
\bibitem [{Note1()}]{Note1}%
  \BibitemOpen
  \bibinfo {note} {See http://www.netlib.org/go/zeroin.f for the original source code of this algorithm}\BibitemShut {NoStop}%
\bibitem [{\citenamefont {Press}\ \emph {et~al.}(2007)\citenamefont {Press},
  \citenamefont {Teukolsky}, \citenamefont {Vetterling},\ and\ \citenamefont
  {Flannery}}]{NumericalRecipes2007}%
  \BibitemOpen
  \bibfield  {author} {\bibinfo {author} {\bibfnamefont {W.~H.}\ \bibnamefont
  {Press}}, \bibinfo {author} {\bibfnamefont {S.~A.}\ \bibnamefont
  {Teukolsky}}, \bibinfo {author} {\bibfnamefont {W.~T.}\ \bibnamefont
  {Vetterling}}, \ and\ \bibinfo {author} {\bibfnamefont {B.~P.}\ \bibnamefont
  {Flannery}},\ }\href@noop {} {\emph {\bibinfo {title} {Numerical Recipes 3rd
  Edition: The Art of Scientific Computing}}},\ \bibinfo {edition} {3rd}\ ed.\
  (\bibinfo  {publisher} {Cambridge University Press},\ \bibinfo {year}
  {2007})\BibitemShut {NoStop}%
\bibitem [{\citenamefont {Shacham}(1986)}]{Shacham1986}%
  \BibitemOpen
  \bibfield  {author} {\bibinfo {author} {\bibfnamefont {M.}~\bibnamefont
  {Shacham}},\ }\href@noop {} {\bibfield  {journal} {\bibinfo  {journal}
  {Internat. J. Numer. Methods Engrg.}\ }\textbf {\bibinfo {volume} {23}},\
  \bibinfo {pages} {1455} (\bibinfo {year} {1986})}\BibitemShut {NoStop}%
\bibitem [{\citenamefont {Berberan-Santos}, \citenamefont {Bodunov},\ and\
  \citenamefont {Pogliani}(2008)}]{BerberanSantos2008}%
  \BibitemOpen
  \bibfield  {author} {\bibinfo {author} {\bibfnamefont {M.}~\bibnamefont
  {Berberan-Santos}}, \bibinfo {author} {\bibfnamefont {E.}~\bibnamefont
  {Bodunov}}, \ and\ \bibinfo {author} {\bibfnamefont {L.}~\bibnamefont
  {Pogliani}},\ }\href {\doibase 10.1007/s10910-007-9272-4} {\bibfield
  {journal} {\bibinfo  {journal} {J. Math. Chem.}\ }\textbf {\bibinfo {volume}
  {43}},\ \bibinfo {pages} {1437} (\bibinfo {year} {2008})}\BibitemShut
  {NoStop}%
\bibitem [{\citenamefont {Younglove}\ and\ \citenamefont
  {McLinden}(1994)}]{Younglove1994}%
  \BibitemOpen
  \bibfield  {author} {\bibinfo {author} {\bibfnamefont {B.~A.}\ \bibnamefont
  {Younglove}}\ and\ \bibinfo {author} {\bibfnamefont {M.~O.}\ \bibnamefont
  {McLinden}},\ }\href@noop {} {\bibfield  {journal} {\bibinfo  {journal} {J.
  Phys. Chem. Ref. Data}\ }\textbf {\bibinfo {volume} {23}},\ \bibinfo {pages}
  {731} (\bibinfo {year} {1994})}\BibitemShut {NoStop}%
\bibitem [{\citenamefont {Wooley}, \citenamefont {Scott},\ and\ \citenamefont
  {Brickwedde}(1948)}]{Wooley1948}%
  \BibitemOpen
  \bibfield  {author} {\bibinfo {author} {\bibfnamefont {H.~W.}\ \bibnamefont
  {Wooley}}, \bibinfo {author} {\bibfnamefont {R.~B.}\ \bibnamefont {Scott}}, \
  and\ \bibinfo {author} {\bibfnamefont {F.~G.}\ \bibnamefont {Brickwedde}},\
  }\href@noop {} {\bibfield  {journal} {\bibinfo  {journal} {J. Res. Natl. Bur.
  Stand.}\ }\textbf {\bibinfo {volume} {41}},\ \bibinfo {pages} {379} (\bibinfo
  {year} {1948})}\BibitemShut {NoStop}%
\bibitem [{\citenamefont {Lemmon}\ and\ \citenamefont
  {Jacobsen}(2005)}]{Lemmon2005}%
  \BibitemOpen
  \bibfield  {author} {\bibinfo {author} {\bibfnamefont {E.~W.}\ \bibnamefont
  {Lemmon}}\ and\ \bibinfo {author} {\bibfnamefont {R.~T.}\ \bibnamefont
  {Jacobsen}},\ }\href {\doibase 10.1063/1.1797813} {\bibfield  {journal}
  {\bibinfo  {journal} {J. Phys. Chem. Ref. Data}\ }\textbf {\bibinfo {volume}
  {34}},\ \bibinfo {pages} {69} (\bibinfo {year} {2005})}\BibitemShut {NoStop}%
\bibitem [{\citenamefont {Baehr}(2005)}]{Baehr2005}%
  \BibitemOpen
  \bibfield  {author} {\bibinfo {author} {\bibfnamefont {H.~D.}\ \bibnamefont
  {Baehr}},\ }\href@noop {} {\emph {\bibinfo {title} {Thermodynamik. Grundlagen
  und technische Anwendungen}}}\ (\bibinfo  {publisher} {Springer-Verlag
  GmbH},\ \bibinfo {year} {2005})\BibitemShut {NoStop}%
\bibitem [{\citenamefont {{VDI-Gesellschaft Verfahrenstechnik und
  Chemieingenieurwesen}}(2010)}]{VDIHeatAtlas}%
  \BibitemOpen
  \bibinfo {editor} {\bibnamefont {{VDI-Gesellschaft Verfahrenstechnik und
  Chemieingenieurwesen}}},\ ed.,\ \href@noop {} {\emph {\bibinfo {title} {VDI
  Heat Atlas}}},\ VDI-Buch\ (\bibinfo  {publisher} {Springer},\ \bibinfo {year}
  {2010})\BibitemShut {NoStop}%
\end{thebibliography}
\end{document}